\documentclass[prb,twocolumn,showpacs,floatfix]{revtex4}

\usepackage{graphicx}
  \graphicspath{{./eps/}}
\usepackage{amsmath}
\usepackage{amssymb}

\begin{document}

\title{Persistent current of Luttinger liquid in one-dimensional ring with weak link:
Continuous model studied by configuration interaction and quantum
Monte Carlo}

\author{R. \surname{N\'emeth}$^{1,2}$}

\author{M. \surname{Mo\v{s}ko}$^{1}$}
\email{martin.mosko@savba.sk}

\author{R. \surname{Kr\v{c}m\'ar}$^1$}

\author{A. \surname{Gendiar}$^1$}

\author{M. \surname{Indlekofer}$^{3}$}

\author{L. Mitas$^4$}

\affiliation{$^1$Institute of Electrical Engineering, Slovak
Academy of Sciences, 841 04 Bratislava, Slovakia}

\affiliation{$^2$Institute for Bio and Nanosystems, CNI, Research
Center J\"{u}lich, 52425 J\"{u}lich, Germany}

\affiliation{$^3$Wiesbaden, University of Applied Sciences,
ING/ITE, 65428 R\"{u}sselsheim, Germany}

\affiliation{$^4$Department of Physics, North Carolina State
University, Raleigh, NC 27695}

\date{\today}

\begin{abstract}
We study the persistent current of correlated spinless electrons
in a continuous one-dimensional ring with a single weak link. We
include correlations by solving the many-body Schrodinger equation
for several tens of electrons interacting via the short-ranged
pair interaction $V(x - x')$. We solve this many-body problem by
advanced configuration-interaction (CI) and diffusion Monte Carlo
(DMC) methods which rely neither on the renormalisation group
techniques nor on the Bosonisation technique of the
Luttinger-liquid model. Our CI and DMC results show, that the
persistent current ($I$) as a function of the ring length ($L$)
exhibits for large $L$ the power law typical of the Luttinger
liquid, $I \propto L^{-1-\alpha}$, where the power $\alpha$
depends only on the electron-electron (e-e) interaction. For
strong e-e interaction the previous theories predicted for
$\alpha$ the formula $\alpha = {(1 + 2 \alpha_{RG})}^{1/2} - 1$,
where $\alpha_{RG} = [V(0)-V(2k_F)]/2\pi \hbar v_F$ is the
renormalisation-group result for weakly interacting electrons,
with $V(q)$ being the Fourier transform of $V(x-x')$. Our
numerical data show that this theoretical result holds in the
continuous model only if the range of $V(x - x')$ is small
(roughly $d \lesssim 1/2k_F$, more precisely $4d^2k_F^2 \ll 1$).
For strong e-e interaction ($\alpha_{RG}\gtrsim 0.3$) our CI data
show the power law $I \propto L^{-1-\alpha}$ already for rings
with only ten electrons, i.e., ten electrons are already enough to
behave like the Luttinger liquid. The DMC data for
$\alpha_{RG}\gtrsim 0.3$ are damaged by the so-called fixed-phase
approximation. Finally, we also treat the e-e interaction in the
Hartree-Fock approximation. We find the exponentially decaying
$I(L)$ instead of the power law, however, the slope of
$\log(I(L))$ still depends solely on the parameter $\alpha_{RG}$
as long as the range of $V(x - x')$ approaches zero.
\end{abstract}


\maketitle


\section{Introduction}

A clean one-dimensional (1D) wire biased by contacts with
negligible backscattering is known to exhibit the conductance
quantized as an integer multiple of $e^2/h$. The effect can be
explained in a Fermi-liquid model of non-interacting
quasi-particles. \cite{Imry-book,Davies-98} In fact, a clean 1D
system is not a Fermi liquid due to the electron-electron (e-e)
interaction. Away from the charge-density-wave instability, the
system is a correlated Luttinger liquid with collective bosonic
elementary excitations which contrast with independent fermionic
quasi-particles of ordinary Fermi liquids \cite{Voit}.
Nevertheless, the Luttinger-liquid model gives for a clean 1D wire
the same conductance ($e^2/h$ per spin) as the Fermi-liquid model
\cite{Maslov}.

If a localized scatterer is introduced into the wire, the
conductance quantization breaks down. For non-interacting
electrons, the Landauer formula \cite{Imry-book,Davies-98}
expresses the conductance per spin as $(e^2/h)|t_{k_{F}}|^2$,
where $t_{k_{F}}$ is the transmission amplitude through the
scatterer at the Fermi level.

In the Luttinger liquid model \cite{Kane,Furusaki}, the infinite
wire which contains a single structureless scatterer, exhibits the
conductance varying with temperature as $\propto T^{2\alpha}$,
where $\alpha$ depends only on the e-e interaction. Thus, for
$\alpha>0$ (repulsive e-e interaction) the wire is impenetrable at
$T \rightarrow 0$ regardless the strength of the scatterer. A
similar power law exhibits at $T \rightarrow 0$ the differential
conductance as a function of the bias voltage ($\propto
U^{2\alpha}$) and conductance versus the wire length ($\propto
L^{-2\alpha}$). These power laws are a sign of the Luttinger
liquid \cite{Kane,Furusaki,Dekker}. The $T^{2\alpha}$ and
$U^{2\alpha}$ laws were successfully measured \cite{Dekker}.

In this work we deal with similar power laws exhibited by
interacting electrons in an isolated mesoscopic 1D ring. In
particular, magnetic flux $\phi$ piercing the opening of the
mesoscopic ring gives rise to the persistent electron current
circulating along the ring \cite{Imry-book}. This persistent
current is given at $T = 0$K as $I= -\partial E_0(\phi)/\partial
\phi$, where $E_0$ is the energy of the many-body groundstate. If
the ring is clean and the single-particle dispersion law is
parabolic, the e-e interaction does not affect the persistent
current due to the Galilean invariance of the system \cite{Shick}.

However, if a single scatterer is introduced, the non-interacting
and interacting result differ fundamentally. For non-interacting
spinless electrons in the 1D ring containing a single scatterer
with transmission probability $|\tilde{t}_{k_{F}}|^2 \ll 1$, the
persistent current at $T = 0$K depends on the magnetic flux and
ring length ($L$) as \cite{Gogolin}
\begin{equation} \label{I-nonint-approx}
I = (ev_F/2 L) \ |\tilde{t}_{k_{F}}| \ \sin(2\pi\phi/\phi_0),
\end{equation}
where $\phi_0 = h/e$ is the flux
quantum, $k_{F}$ is the Fermi wave vector, and $v_F$ is the Fermi
velocity. For a spinless Luttinger liquid the persistent current
follows the power law $I \propto L^{-\alpha-1}$. More precisely,
\cite{Gogolin}
\begin{equation} \label{I-Luttinger}
I \propto L^{-\alpha-1} \sin(2\pi\phi/\phi_0),
\end{equation}
where  the power $\alpha$ depends only on the e-e interaction, not
on the properties of the scatterer. The formulae
\eqref{I-nonint-approx} and \eqref{I-Luttinger} were derived
\cite{Gogolin} assuming large $L$.

The formula \eqref{I-Luttinger} can also be obtained heuristically
\cite{Gogolin} as follows. Matveev et al. \cite{Matveev} analyzed,
how the e-e interaction renormalizes the bare transmission
amplitude $\tilde{t}_{k_{F}}$ of a single scatterer inside the 1D
wire with two contacts. They derived the renormalized amplitude
$t_{k_{F}}$ by using the renormalization-group (RG) approach
suitable for a weakly-interacting electron gas. If the system
length ($L$) is large, their result can be expressed in the form
\begin{equation} \label{transmission-Matveev}
t_{k_{F}} \simeq (\tilde{t}_{k_{F}}/|\tilde{r}_{k_{F}}|)
(d/L)^{\alpha}, \quad \alpha > 0,
\end{equation}
where $|\tilde{r}_{k_{F}}|^2 = 1 - |\tilde{t}_{k_{F}}|^2$, $d$ is
the spatial range of the e-e interaction $V(x-x')$, and the power
$\alpha$ is given (for spinless electrons) by the expression
\begin{equation} \label{power-Matveev}
\alpha_{RG} = [V(0)-V(2k_F)]/2\pi \hbar v_F,
\end{equation}
with $V(q)$ being the Fourier transform of $V(x-x')$. We note that
the formulae \eqref{transmission-Matveev} and
\eqref{power-Matveev} were derived assuming
\begin{equation} \label{weakinteractionMatveev}
1 \ll \ln(l/d) \ll {\alpha_{RG}}^{-1},
\end{equation}
where $l$ is a properly chosen scale of the RG theory
\cite{Matveev}. Moreover, $\alpha=\alpha_{RG}$ only for
$\alpha_{RG} \ll 1$ (weak e-e interaction).  For strong e-e
interaction (say $\alpha_{RG} \simeq 0.5$) the theory
\cite{Voit,Polyakov} predicts the more general result,
\begin{equation} \label{power-Polyakov}
\alpha = {(1 + 2 \alpha_{RG})}^{1/2} - 1.
\end{equation}
This result is believed \cite{Polyakov} to hold for any $V(x-x')$
with range $d$ which is finite but which can in principle be quite
large (in comparison with $1/k_F$). Finally, if we replace in
equation \eqref{I-nonint-approx} the bare amplitude
$\tilde{t}_{k_{F}}$ by the renormalized amplitude
\eqref{transmission-Matveev}, we recover the power law
\eqref{I-Luttinger}, where $\alpha$ is now given by the
microscopic formulae \eqref{power-Polyakov} and
\eqref{power-Matveev}.

 In the Luttinger-liquid model, the physics of the
low-energy excitations is mapped onto an effective field theory
using Bosonization \cite{Gogolin}, where terms expected to be
negligible at low energies are omitted. Within this model, the
asymptotic dependence \eqref{I-Luttinger} was obtained by using
the analogy to the problem of quantum coherence in dissipative
environment \cite{Gogolin}. To avoid this analogy as well as
Bosonization, in Ref. \cite{Meden} the persistent current was
calculated by solving the 1D lattice model with nearest-neighbor
hopping and interaction. Applying numerical RG methods, the
formula \eqref{I-Luttinger} was confirmed for long chains and
strong scatterers. \cite{Meden}

However, insofar it has not been verified whether the RG formula
\eqref{power-Matveev} holds in a microscopic model which does not
rely on the RG approach. We present such microscopic many-body
model in this work. Dealing with a continuous model, we can vary
the range of the e-e interaction in order to test the robustness
of the formula \eqref{power-Matveev} against various shapes of
$V(x - x')$. This point was not addressed in the
lattice-model-based studies as the range of the e-e interaction
was fixed to the nearest-neighbor-site interaction.

Further, derivations \cite{Gogolin,Matveev} of the formulae
\eqref{I-Luttinger}, \eqref{transmission-Matveev} and
\eqref{power-Matveev} rely on the large number of particles limit.
Here we directly address the interesting question what is the
minimum number of electrons which exhibits the onset of the
Luttinger-liquid dependence $I \propto L^{-\alpha-1}$. As we will
illustrate, such behavior can be identified from system sizes of
the order of ten electron.

We study the persistent current of correlated electrons in a
continuous 1D ring containing the strongly-reflecting scatterer,
because strong backscattering is known
\cite{Gogolin,Matveev,Meden} to reduce the system size necessary
to achieve the $L^{-\alpha-1}$ asymptotics. Using advanced
configuration-interaction (CI) and diffusion Monte Carlo (DMC)
methods, we solve the continuous many-body Schrodinger equation
for several tens of electrons interacting via the e-e interaction
\begin{equation} \label{VeeExp}
V(x - x') = V_0 \,  \exp(- \left| x - x' \right|/d).
\end{equation}
Interaction \eqref{VeeExp} emulates screening (say by the metallic
gates) and allows us to compare our results with the results of
the correlated models \cite{Gogolin,Matveev,Meden} which also
assume the e-e interaction of finite range. Our CI and DMC
calculations do not rely on the RG techniques and solve the
continuous model, unlike the RG studies focused largely on the
lattice models \cite{Meden,Enss-2005,Meden-2005,Andregassen}. Our
numerical data show that the formulae \eqref{power-Polyakov} and
\eqref{power-Matveev} hold only if the range of the e-e
interaction is small ($d \lesssim 1/2k_F$). For strong e-e
interaction ($\alpha_{RG}\gtrsim 0.3$) our CI data show the power
law $I \propto L^{-1-\alpha}$ already for rings with only ten
electrons. In other words, ten electrons are already sufficient to
show the Luttinger-liquid behavior.

It is known that the fermion sign problem causes an exponential
inefficiency of the DMC method \cite{Foulkes-01} unless it is
circumvented by the so-called fixed-node or the fixed-phase
approximation for inherently complex wave functions \cite{Ortiz}.
The accuracy of our DMC results for $\alpha_{RG}\gtrsim 0.3$ is
therefore limited by the quality of the phase in the Hartree-Fock
trial wave function which is employed in the fixed-phase
approximation. Since Hartree-Fock is the simplest possible trial
wave function and does not capture the Luttinger-liquid
correlation effects it is not surprising that the fixed-phase bias
becomes pronounced. We analyze these findings later in detail.
(Our DMC should not be confused with the path-integral Monte-Carlo
methods \cite{Fisher-93,Egger-93,Egger-95,Egger-2004}, used to
study the tunnelling conductance of the Luttinger liquid in the
bosonised model.)

Finally, we treat the e-e interaction in the self-consistent
Hartree-Fock calculation. We find for large $L$ the exponentially
decaying $I(L)$ instead of the power law. However, the slope of
$\log(I(L))$ still depends solely on the parameter $\alpha_{RG}$
given by the RG formula \eqref{power-Matveev}, as long as the
range of $V(x - x')$ approaches zero.

In section II.A we start with the single-particle approach to the
1D ring. In section II.B we define our interacting many-body
model. In section II.C we outline how we solve the many-body model
in the Hartree-Fock approximation. In sections II.D and II.E we
describe how we obtain the fully-correlated many-body solution by
means of the DMC method and CI method. Our results are discussed
in section III and a summary is given in section IV. Technical
details are in the appendices.

\section{Theory}

\subsection{Single-particle model}

We consider the circular 1D ring threaded by magnetic flux
$\phi=BS=AL$, where $S$ is the area of the ring, $B$ is the
magnetic field (constant and perpendicular to the ring area), and
$A$ is the magnitude of the resulting vector potential
(circulating along the ring circumference). In this section we
discuss the single-particle states in such ring. In general, the
single-electron wave functions $\psi_n(x)$ in the 1D ring obey the
Sch\"{o}dinger equation
\begin{equation}\label{independent Eqs-Schroding-1}
  \left[\frac{\hbar^2}{2m}\left(\frac{1}{i}\frac{\partial}{\partial x}+
    \frac{2\pi}{L}\frac{\phi}{\phi_0}\right)^2 + U(x) \right]\psi_n(x)=\varepsilon_n\psi_n(x)
\end{equation}
with the cyclic boundary condition
\begin{equation} \label{cyclic Eqs-Boundary}
  \psi_n(x+L)=\psi_n(x)\,,
\end{equation}
where $m$ is the electron effective mass, $x$ is the electron
coordinate along the ring, and $U(x)$ is an external
single-particle potential.

We introduce the wave functions $\varphi_n(x)$ by substitution
\begin{equation}\label{independent_Eqs-Substit}
  \psi_n(x)=\varphi_n(x)\exp{\left(-i\frac{2\pi}{L}\frac{\phi}{\phi_0}x\right)}\,.
\end{equation}
If we set \eqref{independent_Eqs-Substit} into \eqref{independent
Eqs-Schroding-1} and \eqref{cyclic Eqs-Boundary}, we obtain the
equation
\begin{equation}\label{independent Eqs-Schroding-2}
  \left[-\frac{\hbar^2}{2m}\frac{d^2}{dx^2} + U(x) \right]\varphi_n(x)=\varepsilon_n\varphi_n(x)
\end{equation}
with the boundary condition
\begin{equation}\label{independent Eqs-Boundary}
  \varphi_n(x+L)=\exp\left(i2\pi\frac{\phi}{\phi_0}\right)\varphi_n(x)\,.
\end{equation}

Equations \eqref{independent Eqs-Schroding-2} and
\eqref{independent Eqs-Boundary} can be solved for an arbitrary
potential $U(x)$ numerically. Consider first the ring region $x
\in \langle-L/2,L/2\rangle$ as a straight-line segment of an
infinite 1D wire. Inside the segment the potential is $U(x)$,
outside we keep it zero. Therefore, the wave function outside is
\begin{equation}\label{Eqs-Leftwave}
  \varphi_k(x)=ae^{ikx}+be^{-ikx}\,,\; x \leq -L/2\,,
\end{equation}
\begin{equation}\label{Eqs-Rightwave}
  \varphi_k(x)=ce^{ikx}+de^{-ikx}\,,\; x \geq L/2\,.
\end{equation}
The amplitudes $a$ and $b$ are related to $c$ and $d$ by
\begin{equation}\label{relationabcd}
  \left(\begin{array}{c}
      c \\ d
      \end{array}
  \right)=T_0
  \left(\begin{array}{c}
      a \\ b
      \end{array}
  \right)\,,
\end{equation}
where $T_0$ is the transfer matrix \cite{Davies-98}
\begin{equation}\label{transfermatrix}
  T_0=\left(
      \begin{array}{cc}
      \frac{1}{t_k^*} & -\frac{r_k^*}{t_k^*} \\
      -\frac{r_k}{t_k} & \frac{1}{t_k}
      \end{array}
  \right)\,,
\end{equation}
with $t_k$ and $r_k$ being the transmission and reflection
amplitudes of the electron impinging the region
$\langle-L/2,L/2\rangle$ from the left. At the boundaries we
express $\varphi_k(-L/2)$ and $\varphi_k(L/2)$ by using equations
\eqref{Eqs-Leftwave} and \eqref{Eqs-Rightwave}. To come back to
the ring threaded by magnetic flux, we relate $\varphi_k(-L/2)$
and $\varphi_k(L/2)$ through the boundary condition
\eqref{independent Eqs-Boundary}. Combining the obtained relation
with the equations (\ref{relationabcd}) and (\ref{transfermatrix})
we obtain the equation
\begin{equation}\label{eigenvalueeq}
  T
  \left(\begin{array}{c}
      a \\ b
      \end{array}
  \right)=\exp(i2\pi\frac{\phi}{\phi_0})
  \left(\begin{array}{c}
      a \\ b
      \end{array}
  \right)\,,
\end{equation}
where
\begin{equation}\label{transfermatrixshifted}
  T=\left(
      \begin{array}{cc}
      \frac{1}{t_k^*}e^{ikL} & -\frac{r_k^*}{t_k^*}e^{ikL} \\
      -\frac{r_k}{t_k}e^{-ikL} & \frac{1}{t_k}e^{-ikL}
      \end{array}
  \right)\,.
\end{equation}
Thus $\exp(i2\pi\phi/\phi_0)$ is the eigenvalue of the matrix $T$.
The product of the eigenvalues of this matrix is given by its
determinant which is unity. The second eigenvalue is thus
$\exp(-i2\pi\phi/\phi_0)$. Their sum is equal to the matrix trace
\cite{Davies-98} , which gives the equation for the spectrum
\cite{remark1},
\begin{equation}\label{Eqs-trancedent}
  \cos\left(2\pi\frac{\phi}{\phi_0}\right)={\rm
  Re}\left[\frac{\exp(-ikL)}{t_k}\right]\,.
\end{equation}
The numerical solution  of equation \eqref{Eqs-trancedent} has to
be combined with numerical computation of the transmission
amplitude $t_k$ (the algorithm for computation of $t_k$ and $r_k$
is described in the Appendix A). The solution of equation
\eqref{Eqs-trancedent} gives us the dependence $k_n(\phi)$ and
eventually the single-particle eigenenergy
$\varepsilon_n(\phi)=\hbar^2k_n^2(\phi)/2m$.

For each $k_n(\phi)$ we can also calculate the wave function
$\varphi_n(x)$. We proceed as follows. By means of equations
\eqref{eigenvalueeq} and \eqref{transfermatrixshifted} we express
the amplitude $a$ in the form
\begin{equation}
  a=\left[\frac{1}{r_{k_n}}-\frac{t_{k_n}}{r_{k_n}}e^{i(2\pi\phi/\phi_0+k_nL)}\right]b\,,
\end{equation}
where the amplitude $b$ can be obtained by normalizing the wave
function. Then we express from equation \eqref{Eqs-Leftwave} the
boundary conditions $\varphi_n(-L/2)$ and $d\varphi_n(-L/2)/dx$.
Finally, we combine these boundary conditions with numerical
solution of equation \eqref{independent Eqs-Schroding-2} in the
discrete form \cite{Press}
\begin{multline} \label{Stoermer}
\varphi_n \left( x_{j+1} \right) = \\
\left[ 2 + \frac {2m} {\hbar^2} \left( U \left(x_j\right) -
\varepsilon_n  \right) \Delta^2 \right] \varphi_n \left(x_j\right)
- \varphi_n \left(x_{j-1}\right) \, ,
\end{multline}
where $\Delta$ is the step, $x_j=j \Delta$, and $j=0, \pm1, \pm2,
\dots$. Once the wave functions $\varphi_n(x)$ are known, the wave
functions $\psi_n(x)$ can be obtained by means of the relation
\eqref{independent_Eqs-Substit}.

\subsection{Interacting many-body model}

We now consider the ring with $N$ interacting 1D electrons. This
system is described by the Hamiltonian
\begin{eqnarray}\label{Eqs-Hamilt-1}
 \nonumber
 \hat{H} &=& \sum\limits _{j=1}^N\left[\frac{\hbar^2}{2m}\left(
    \frac{1}{i}\frac{\partial}{\partial x_j}
    +\frac{2\pi}{L}\frac{\phi}{\phi_0}\right)^2
    +\gamma\delta(x_j)\right] \\
   &+&\frac{1}{2}\sum \limits_{{i,j=1}\atop{i\neq j}}^N V(x_j-x_i)\,,
\end{eqnarray}
where $x_j$ is the coordinate of the $j$-th electron,
$\gamma\delta(x)$ is the potential of the scatterer, and
$V(x_j-x_i)$ is the e-e interaction \eqref{VeeExp}. The
eigenfunction $\Psi(x_1,x_2,\dots,x_N)$ and eigenenergy $E$ obey
the Sch\"{o}dinger equation
\begin{equation} \label{originalmany-body}
  \hat{H}\Psi=E\Psi
\end{equation}
with the cyclic boundary condition
\begin{equation} \label{cyclicboundarycondition}
  \Psi(x_1,\dots,x_i+L,\dots,x_N)=\Psi(x_1,\dots,x_i,\dots,x_N)
\end{equation}
for $i=1,2,\dots,N$.

If the system is in the groundstate with eigenfunction
$\Psi_0(x_1,x_2,\dots,x_N)$ and eigenenergy $E_0$, the persistent
current can be expressed as
\begin{equation} \label{stred-Operator-Prud}
I = \left< \Psi_0 \right| \hat{I} \left| \Psi_0 \right>\,
\end{equation}
where
\begin{equation} \label{Operator-Prud}
\hat{I} = -\frac{e}{mL} \sum\limits_{j=1}^N
\left(\frac{\hbar}{i}\frac{\partial}{\partial x_j} + \frac{e
\phi}{L}\right)
\end{equation}
is the $N$-particle current operator. One can calculate $I$
directly from the relation \eqref{stred-Operator-Prud}, or one can
rewrite \eqref{stred-Operator-Prud} by means of the
Hellman-Feynman theorem
\begin{equation} \label{Hellman-Feynman}
\frac{\partial E}{\partial \phi} = \left< \Psi_0 \right|
\frac{\partial \hat{H}}{\partial \phi} \left| \Psi_0 \right>\, .
\end{equation}
Using \eqref{Hellman-Feynman}, \eqref{Operator-Prud}, and
\eqref{Eqs-Hamilt-1} one gets the formula \cite{Imry-book,Eckern}
\begin{equation} \label{Eqs-Vseob-Prud}
  I= -\frac{\partial}{\partial \phi}E_0(\phi)\,.
\end{equation}

\subsection{Hartree-Fock approximation}

In the Hartree-Fock model, the ground-state wave function $\Psi_0$
is approximated by the Slater determinant
\begin{equation} \label{SlaterHFground}
  \Psi_0(x_1,\dots,x_N)=
  \frac{1}{\sqrt{N!}}\left|\begin{array}{ccc}
                          \psi_{1}(x_1)&\cdots&\psi_{1}(x_N)\\
                \vdots& \ddots&\vdots\\
                \psi_{N}(x_1)&\cdots&\psi_{N}(x_N)
                   \end{array}\right|\,.
\end{equation}
The wave functions $\psi_n(x)$ obey the Hartree-Fock equation
\begin{multline}\label{Eqs-Schroding-1}
  \left[\frac{\hbar^2}{2m}\left(-i\frac{\partial}{\partial x}+
    \frac{2\pi}{L}\frac{\phi}{\phi_0}\right)^2+\gamma\delta(x)\right.\\
    \left.\phantom{\frac{\hbar^2}{2m}}+U_H(x)+U_F(n,x)\right]\psi_n(x)=\varepsilon_n\psi_n(x)
\end{multline}
with the boundary condition
\begin{equation}
  \psi_n(x+L)=\psi_n(x)\,,
\end{equation}
where
\begin{equation}\label{Eqs-Hartree}
  U_H(x)=\sum_{n'} \int dx' V(x-x')|\psi_{n'}(x')|^2
\end{equation}
is the Hartree potential and
\begin{multline}\label{Eqs-Fock}
  U_F(n,x)=\\
    -\frac{1}{\psi_n(x)}\sum \limits_{n'}
        \int dx' V(x-x')\psi_n(x')\psi_{n'}^*(x')\psi_{n'}(x)
\end{multline}
is the Fock nonlocal exchange term (expressed as an effective
potencial for further convenience). In the equations
\eqref{Eqs-Hartree} and \eqref{Eqs-Fock} we sum over all occupied
states $n'$.

If we use again the substitution
\begin{equation}\label{Eqs-Substit}
  \psi_n(x)=\varphi_n(x)\exp{\left(-i\frac{2\pi}{L}\frac{\phi}{\phi_0}x\right)}\,,
\end{equation}
the equations \eqref{Eqs-Schroding-1}-\eqref{Eqs-Fock} give the
Hartree-Fock equation
\begin{multline}\label{Eqs-Schroding-2}
  \left[-\frac{\hbar^2}{2m}\frac{d^2}{dx^2}+\gamma\delta(x)+U_H(x)+U_F(n,x)\right]\varphi_n(x)\\
  =\varepsilon_n\varphi_n(x)
\end{multline}
with the boundary condition
\begin{equation}\label{HF Eqs-Boundary}
  \varphi_n(x+L)=\exp\left(i2\pi\frac{\phi}{\phi_0}\right)\varphi_n(x)\,,
\end{equation}
where the potentials $U_H(x)$ and $U_F(n,x)$ are still given by
equations \eqref{Eqs-Hartree} and \eqref{Eqs-Fock}, but with
$\psi_n$ replaced by $\varphi_n$. From equations
\eqref{Eqs-Hamilt-1}-\eqref{HF Eqs-Boundary} the groundstate
energy $E_0=\langle\Psi_0|H|\Psi_0\rangle$ can be expressed as
\begin{equation} \label{HF ground state energy}
  E_0=\sum \limits _n
  \left[\varepsilon_n-\frac{1}{2}\left\langle\varphi_n\left|U_H(x)+U_F(n,x)\right|\varphi_n\right\rangle\right]\,.
\end{equation}

The Hartree-Fock equation \eqref{Eqs-Schroding-2} can be solved by
the same procedure as the single-particle equation
\eqref{independent Eqs-Schroding-2} assuming that the potential
$U(x) \equiv \gamma\delta(x)+U_H(x)+U_F(n,x)$ is known. We apply
this procedure iteratively in order to obtain the self-consistent
Hartree-Fock solution. In the first iteration step we solve
equation \eqref{Eqs-Schroding-2} for the non-interacting gas,
i.e., for $U_H(x)=0$ and $U_F(n,x)=0$. The resulting
$\varphi_n(x)$ is used to evaluate the Hartree and Fock
potentials, where the term $U_F(n,x)$ has to be evaluated for each
$n$ separately. These potentials are used in the second iteration
step to obtain new $\varphi_n(x)$ and new potentials $U_H(x)$ and
$U_F(n,x)$, etc., until the energies $\varepsilon_n$ and
ground-state energy \eqref{HF ground state
 energy} do not change anymore. In the Appendix B the iteration procedure is described
 including a few nontrivial details. Setting the resulting ground-state energy
\eqref{HF ground state energy} into \eqref{Eqs-Vseob-Prud} we
obtain the persistent current. Of course, this Hartree-Fock
calculation does not include the many-body correlations. To
include the correlations we use the DMC and CI techniques
described in the next two sections.

\subsection{Diffusion Monte Carlo (DMC) model}

Consider first the $N$-electron 1D Sch\"{o}dinger equation
\begin{equation} \label{SchrDM}
\hat{\mathcal{H}}\Psi(\textbf{X})=E\Psi(\textbf{X})
\end{equation}
with Hamiltonian
\begin{equation} \label{HamiltDM}
\hat{\mathcal{H}}=
-\frac{\hbar^2}{2m}\sum_{i=1}^N\frac{\partial^2}{\partial x_i^2} +
V(X),
\end{equation}
where  $\textbf{X}=(x_1, x_2, \dots, x_N)$, the potential energy
$V(X)$ incorporates all single-electron interactions and all pair
electron-electron interactions, and the wave-function
$\Psi(\textbf{X})$ is assumed to obey the cyclic condition
\eqref{cyclicboundarycondition}. If $\Psi(\textbf{X})$ is a real
function, the DMC method is capable to find the exact ground-state
solution of equation \eqref{SchrDM}. We briefly summarize how the
DMC works \cite{Foulkes-01}.

Instead of directly solving the equation \eqref{SchrDM}, the DMC
solves the time-dependent diffusion problem \cite{Foulkes-01}
\begin{equation} \label{imaginarytimeSchr}
-\partial \Psi(\textbf{X},t)/
\partial t = \left(\hat{\mathcal{H}}-E_T \right) \Psi(\textbf{X},t),
\end{equation}
where $E_T$ is a trial energy. One can write
\eqref{imaginarytimeSchr} in the integral form \cite{Foulkes-01}
\begin{equation} \label{GF} \Psi(\textbf{X},t+\tau)= \int d\textbf{X'} \
G_{\textbf{X} \leftarrow \textbf{X'}}(\tau) \ \Psi(\textbf{X'},t),
\end{equation}
where $G_{\textbf{X} \leftarrow \textbf{X'}}(\tau) = \langle
\textbf{X} | \exp (-(\hat{\mathcal{H}}-E_T) \tau ) | \textbf{X'}
\rangle$ is the Green's function and $\tau$ is a time step.
Equation \eqref{GF} allows to project the ground state
$\Psi_0(\textbf{X})$ from any trial wave function
$\Psi_T(\textbf{X})$ which has a nonzero overlap with
$\Psi_0(\textbf{X})$. Inserting any trial $\Psi_T(\textbf{X})$ and
$E_T$ into \eqref{GF} gives
\begin{equation} \label{trialinsert}
\Psi_T (\textbf{X}, \tau \rightarrow \infty ) = \Psi_0(\textbf{X})
\langle \Psi_0|\Psi_T \rangle \lim \limits _{\tau \rightarrow
\infty} \exp[-\tau(E_0-E_T) ].
\end{equation}
By adjusting $E_T$ to equal $E_0$ one can make the exponential
factor constant. It is also important that the Green's function
$G_{\textbf{X} \leftarrow \textbf{X'}}(\tau)$ can be expressed
\cite{Foulkes-01} for $\tau \rightarrow 0$ as
\begin{multline} \label{greenovafunkcia}
     G_{X\leftarrow X'}(\tau) \approx
         (2\pi\tau)^{-3N/2}\exp\left[-\frac{(X-X')^2}{2\tau}\right]
\\
\times\exp\left(-\tau\left[V(X)+V(X')-2E_T\right]/2\right) .
\end{multline}

In the DMC algorithm the equation \eqref{GF} is solved
stochastically based on the action of the projection operator
$\exp[-\tau (\hat{\mathcal{H}}-E_T)]$ \cite{Foulkes-01}. At time
$t=0$ one chooses a proper trial wave function
$\Psi_T(\textbf{X})$. A set of walkers (sampling points in the
$N$-dimensional electron configuration space) is generated
according to the distribution $\Psi_T(\textbf{X})$. A small
timestep $\tau$ is chosen. The walkers are propagated from
$\textbf{X}$ to $\textbf{X'}$ according to the kernel
$G_{\textbf{X} \leftarrow \textbf{X'}}(\tau)$ in the form
\eqref{greenovafunkcia}. In the new sample point $\textbf{X'}$,
the expectation values are calculated and the starting value of
$E_T$ is adjusted. Eventually, the ground state expectation values
are projected and sampled.

The problem is that the wave function $\Psi(\textbf{X})$ obeying
the many-body equation \eqref{originalmany-body} is complex
because the Hamiltonian \eqref{Eqs-Hamilt-1} contains the magnetic
flux. The problem can be partly eliminated by means of the
fixed-phase approximation \cite{Ortiz}. Using
\begin{equation} \label{trialfixedphase}
\Psi(\textbf{X}) = \left| \Psi(\textbf{X}) \right|
e^{i\Phi(\textbf{X})}
\end{equation}
one can split the equation \eqref{originalmany-body} into the real
part and imaginary part. The real part reads
\begin{equation} \label{fixedphaseapprox}
 \hat{H}_{\text{eff}} \left|
\Psi(\textbf{X}) \right| = E \left| \Psi(\textbf{X}) \right|,
\end{equation}
where
\begin{eqnarray} \label{effectiveHamiltonian}
\nonumber \hat{H}_{\text{eff}}=
-\frac{\hbar^2}{2m}\sum_{i=1}^N\frac{\partial^2}{\partial x_i^2} +
\frac{1}{2m}\sum_{i=1}^N\left(\hbar\frac \partial{\partial
x_i}\Phi +
 \frac{e\phi}{L}\right)^2\\
 +\gamma\sum_{i=1}^N\delta(x_i) + \frac12
 \sum\limits_{{i,j=1}\atop{i\neq j}}^NV(x_i-x_j),
\end{eqnarray}
and the imaginary part is
\begin{equation} \label{eqautiomforphase}
 \sum_{i=1}^N\frac \partial{\partial x_i}\left[|\Psi|^2\left(\hbar\frac \partial{\partial x_i}\Phi +
\frac{e\phi}{L}\right)\right] = 0.
\end{equation}
Equation \eqref{fixedphaseapprox} is the effective Sch\"{o}dinger
equation for the modul $\left| \Psi(\textbf{X}) \right|$, with the
Hamiltonian $\hat{H}_{\text{eff}}$ depending on the phase
$\Phi(\textbf{X})$. Since $\left| \Psi(\textbf{X}) \right|$ is
real and $\hat{H}_{\text{eff}}$ has the same form as
\eqref{HamiltDM}, the equation \eqref{fixedphaseapprox} can be
solved by means of the DMC if the phase $\Phi(\textbf{X})$ is
given. In this work we choose the trial wave function
$\Psi_T(\textbf{X})$ to be equal to the Slater determinant
\eqref{SlaterHFground} of the self-consistently determined
Hartree-Fock ground state and we fix the phase $\Phi(\textbf{X})$
to the phase of this Slater determinant. The DMC thus gives the
lowest possible ground-state energy $E_0$ within the chosen phase
\cite{Ortiz}. Eventually, we obtain the persistent current from
\eqref{Eqs-Vseob-Prud} by finite differences evaluated by
correlated sampling \cite{Filipi}. Our preliminary DMC results are
briefly discussed in \cite{Vagner,Krcmar}.

To go beyond the fixed phase approximation one has to couple the
DMC solution of equation \eqref{fixedphaseapprox} with the
numerical solution of equation \eqref{eqautiomforphase}. We do not
attempt to do so. However, in section III we check the fixed phase
approximation by comparing the DMC results with the CI
calculations which are free of such approximation.

\subsection{Configuration-interaction (CI) model}

Before starting with the CI procedure we need to solve the
single-electron problem
\begin{equation}\label{1-e}
\left[\frac{\hbar^2}{2m}\left(\frac{1}{i}\frac{\partial}{\partial
x}+
    \frac{2\pi}{L}\frac{\phi}{\phi_0}\right)^2 + \gamma\delta(x)
    \right]\psi_n(x)=\varepsilon_n\psi_n(x)
\end{equation}
with the cyclic condition $\psi_n(x+L)=\psi_n(x)$. We do so by
using the method described in Section II.A. We obtain the wave
functions $\psi_n(x)$ and energy levels $\varepsilon_n$ not only
for the $N$ lowest energy levels but also for the infinite ladder
of excited states, of course, with an upper cutoff.

\begin{figure}[t]
\centerline{\includegraphics[clip,width=\columnwidth]{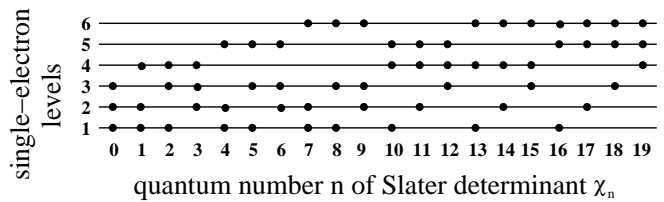}}
\caption{Schematic sketch of how the determinants $\chi_n$ are
created for $n = 0, 1, \dots$ in the three-electron system with
the ladder of single-particle levels restricted to six lowest
levels. The occupied levels constituting the state $\chi_n$ are
labelled by circles. In the CI method the many-body state $\Psi$
of the interacting system is expanded as $ \Psi = c_0\chi_0 +
c_1\chi_1 + c_2\chi_2 + \dots$.} \label{Fig:CI}
\end{figure}

Consider now the non-interacting many-body problem
\begin{equation}\label{MBS}
 \sum_{j=1}^N\left(
\frac{\hbar^2}{2m}\left(\frac{1}{i}\frac{\partial}{\partial x_j}+
    \frac{2\pi}{L}\frac{\phi}{\phi_0}\right)^2
 +
\gamma\delta(x_j)\!\!\right) \chi_n(\textbf{X}) =
\mathcal{E}_n\chi_n(\textbf{X}),
\end{equation}
where $\textbf{X}=(x_1, x_2, \dots, x_N)$. Clearly, the
eigenenergies $\mathcal{E}_n$ are given as
\begin{equation}\label{MBS-energy}
\mathcal{E}_n=\varepsilon_{n_1}+\dots+\varepsilon_{n_N}
\end{equation}
and the wave functions $\chi_n$ are the Slater determinants
\begin{equation}\label{SLater}
 \chi_n = \frac{1}{\sqrt{N!}}
        \left|
        \begin{array}{ccc}
        \psi_{n_1}(x_1) & \dots & \psi_{n_N}(x_1)\\
        \vdots & \ddots & \vdots\\
        \psi_{n_1}(x_N) & \dots & \psi_{n_N}(x_N)
\end{array}\right|,
\end{equation}
where the quantum numbers $n_1, \dots, n_N$ label the state of the
first, $\dots, N$-th electron, respectively, and the quantum
number $n$ labels the many-body state corresponding to the
specific set of $N$ occupied single-electron levels
$\varepsilon_{n_1},\dots,\varepsilon_{n_N}$. For instance, the
figure \ref{Fig:CI} depicts creation of all Slater determinants in
the three-electron system, when only six lowest single-electron
levels are considered.

To solve the interacting many-body problem
\eqref{originalmany-body}, the CI method \cite{Szabo,Jensen}
relies on the expansion
\begin{equation}\label{SLaterexpansion}
 \Psi = c_0\chi_0 + c_1\chi_1 + c_2\chi_2 + \dots \\.
\end{equation}
Using this expansion and equation
$\langle\chi_n|\hat{H}|\Psi\rangle = \langle\chi_n|E|\Psi\rangle$
we obtain from \eqref{originalmany-body} the infinite set of
equations
\begin{equation} \label{infiniteset}
\sum_{j=0}^\infty\left(\mathcal{E}_j\delta_{nj} + V_{nj}\right)c_j
= Ec_n, \quad  n = 0,1, \dots, \infty,
\end{equation}
where
\begin{equation} \label{matrixelementVij}
V_{nj} = \frac{1}{2}\langle\chi_n|\sum\limits_{{k,l=1}\atop{k\neq
l}}^{N} V(x_k - x_l)|\chi_j\rangle.
\end{equation}

We reduce the infinite number of single-energy levels
$\varepsilon_j$ to the finite one by introducing a proper upper
energy cutoff. This reduces the infinite number of equations
\eqref{infiniteset} to a certain finite number $M+1$. We get the
finite system
\begin{equation} \label{infiniteset1}
\sum_{j=0}^M\left(\mathcal{E}_j\delta_{nj} + V_{nj}\right)c_j =
Ec_n,
\end{equation}
where $n = 0, 1, \dots, M$. This system determines the eigenvalues
$E_l$ and eigenvectors $(c^{l}_0, c^{l}_1, \dots, c^{l}_M)$ for $l
= 0, 1, \dots, M$. We obtain the groundstate energy $E_{l=0}$ and
groundstate wave function $\Psi_{l=0}$ by solving the system with
ARPACK, or alternatively with LAPACK. Once $E_{0}$ and
$\Psi_{0}$ are known, the persistent current can be obtained from
\eqref{Eqs-Vseob-Prud} or \eqref{stred-Operator-Prud}.

The problem of the CI method is, that for $N_{max}$
single-particle levels the expansion \eqref{SLaterexpansion} still
contains $\left({N_{max}}\atop{N}\right)$ Slater determinants. For
a reasonably chosen cutoff, the solution of the equations
\eqref{infiniteset1} is not feasible already for systems with
about ten electrons, as the memory requirements are too large. The
question is which determinants have to be kept in the expansion
\eqref{SLaterexpansion} and which can be omitted without damaging
the final results. This problem is known in the quantum chemistry
\cite{Szabo,Jensen} where the CI is applied to calculate the
energy spectra of molecules. We cannot apply here directly the
tricks from the quantum chemistry because our problem is rather
special; we search for the persistent current in the 1D ring
pierced by magnetic flux. We prefer the following two approaches.

Our first CI approach, below referred as the FCI, conceptually
resembles the full CI and can be understood by means of the sketch
in the figure \ref{Fig:CI}. After choosing the cutoff for the
single-particle energy we add into the expansion
\eqref{SLaterexpansion} first the Slater determinant of the ground
state, then all determinants with a single excited electron (the
single-excitations), all determinants with two excited electrons
(the double-excitations), all determinants with three excited
electrons (the triple-excitations), etc. If the results do not
change, we cut the expansion by stopping to increase the number of
the excited electrons. Some preliminary FCI results are briefly
discussed in \cite{Krcmar}.

Our second CI approach, introduced by two of us in Ref.
\cite{Indlekofer}, is called the bucket-brigade CI method (BBCI).
After choosing the cutoff for the single-particle energy, we need
to asses importance of the remaining Slater determinants in the
expansion \eqref{SLaterexpansion}. In principle, one can compute
for each determinant $\chi_n$ the energy $\left< \chi_n \right|
\hat{H} \left| \chi_n \right>$, where $\hat{H}$ is the Hamiltonian
\eqref{Eqs-Hamilt-1}. The obtained numerical values $\left< \chi_n
\right| \hat{H} \left| \chi_n \right>$ can be ordered increasingly
and we can assume that this orders the determinants $\chi_n$
according to their decreasing importance. We can order the
determinants in the expansion \eqref{SLaterexpansion} according to
this importance criterion (a-priori measure of importance) and we
can truncate the expansion after reaching convergence of the final
results. This is the basic idea of the BBCI method, the term
"bucket brigade" is related to the technical implementation
\cite{Indlekofer} reviewed in the Appendix C.

In the next section, reliability of our importance criterion is
supported by a successful convergence of the BBCI results. We
cannot prove the criterion exactly, but we can give a simple
intuitive motivation. We search for the best minimum of the
ground-state energy. Therefore, the larger the energy $\left<
\chi_n \right| \hat{H} \left| \chi_n \right>$ the smaller should
be the weight of the given $\chi_n$ in the ground-state expansion
\eqref{SLaterexpansion}. It is important that the BBCI allows us
to treat more electrons than the FCI, as it reduces the expansion
\eqref{SLaterexpansion} more efficiently. Another CI algorithm,
more efficient than the FCI, is the CI approach of Ref.
\onlinecite{Greer}, where the determinants are selected by means
of a proper Monte Carlo algorithm.



\section{Results}

\subsection{Luttinger-liquid scaling in CI and DMC models}

Our calculations are performed for parameters typical of a GaAs
ring. We use the electron effective mass $m = 0.067m_0$ and
electron density $n_e = N/L=5 \times 10^7$ m$^{-1}$. The strength
$V_0$ and range $d$ of our e-e interaction \eqref{VeeExp} are
properly varied to demonstrate various e-e interaction effects. In
practice, screening by metallic gates can be used to vary $d$
while $V_0$ can be varied by varying the (finite) cross-section of
the 1D ring. We study the rings containing a strong scatterer with
$|\tilde{t}_{k_{F}}|^2 \ll 1$, because in such case the persistent
current reaches asymptotic dependence on $L$ for relatively small
$L$. Our conclusions hold for any 1D system, no matter what
material is used.

\begin{figure}[t]
\centerline{\includegraphics[clip,width=\columnwidth]{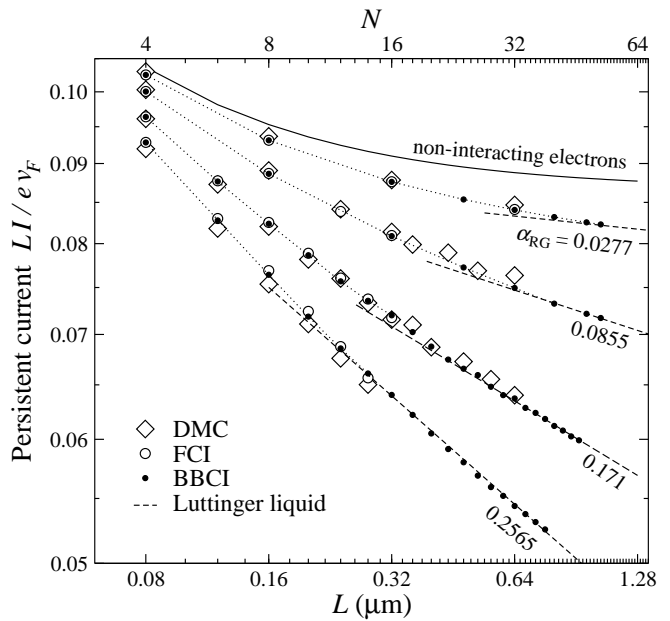}}
\vspace{-0.15cm} \caption{Persistent current $LI/ev_F$ versus the
ring length $L$ for the GaAs ring with a single scatterer. The
transmission of the scatterer is $|\tilde{t}_{k_{F}}|^2 = 0.03$,
magnetic flux is $\phi = 0.25\phi_0$. The upper horizontal axis
shows the electron number $N = n_e L$, where $n_e=5 \times 10^7$
m$^{-1}$. The range of the e-e interaction is fixed to $d = 3$nm
while the magnitude $V_0$ is varied. The full curve is our
numerical result for $V_0 = 0$ (non-interacting electrons). This
curve saturates for large $L$ exactly at the value
$0.5|\tilde{t}_{k_{F}}|$, predicted by the asymptotic formula
\eqref{I-nonint-approx}. The symbols show our FCI, BBCI and DMC
results for various $V_0$ shown in table \ref{Tab-alpha}, the
dotted lines connecting the BBCI data are a guide for eye. The
dashed lines show the Luttinger liquid asymptotics $LI \ \propto
L^{-\alpha}$ for $\alpha = {(1 + 2 \alpha_{RG})}^{1/2} - 1$, where
the values of $\alpha_{RG}$ (listed in the figure) are calculated
from the formula \eqref{power-Matveev2}. The input parameters
$V_0$ and $d$ and the resulting $\alpha_{RG}$ and $\alpha$ are
summarized in the table \ref{Tab-alpha}.} \label{Fig:2}
\end{figure}

In figure \ref{Fig:2} the persistent current $I$ in the GaAs ring
with one strong scatterer is calculated as a function of the ring
length $L$ for magnetic flux $\phi = 0.25\phi_0$. The scatterer is
the $\delta$ barrier with transmission $|\tilde{t}_{k_{F}}|^2 =
0.03$. The power law $LI \propto L^{-\alpha}$ decays in the log
scale linearly with slope $-\alpha$, the same decay show for large
$L$ the CI and DMC data. Agreement of the FCI and BBCI data is
very good, the DMC data are close to the CI data.

Let us check whether our numerical values of $\alpha$ agree with
the formula \eqref{power-Polyakov}. The Fourier transform of our
e-e interaction \eqref{VeeExp} is $V(q) = 2V_0 d/(1+q^2d^2)$.
Using this expression we can write the RG formula
\eqref{power-Matveev} in the form
\begin{equation} \label{power-Matveev2}
\alpha_{RG} = \frac{4 V_0 m k_F d^3}{\pi \hbar^2 (1+4{k_F}^2d^2)}.
\end{equation}
The dashed lines in the figure \ref{Fig:2} show the power law $LI
\propto L^{-\alpha}$, with $\alpha$ given by the formula
\eqref{power-Polyakov} and $\alpha_{RG}$ calculated from the
formula \eqref{power-Matveev2}. The proportionality factor
(specified later on) causes in the log scale only a vertical shift
of the linear law $\log(LI) = -\alpha \log(L)+$\emph{const}. We
thus can conclude that the CI and DMC data in the figure
\ref{Fig:2} exhibit the $L^{-\alpha}$ decay with the power
$\alpha$ in good agreement with the theoretical formulae
\eqref{power-Polyakov} and \eqref{power-Matveev2}.
\begin{table} [b]
    \begin{center}
        \begin{tabular}{|c|c||c|c|}
            \hline
            $V_0$ (meV) &   $d$ (nm)    &   $\alpha_{RG}$   &   $\alpha$    \\
            \hline
            \hline
                11      &       3       &       0.0277      &   0.0273      \\
                34      &       3       &       0.0855      &   0.0821      \\
                68      &       3       &       0.171       &   0.158       \\
                102     &       3       &       0.2565      &   0.230       \\
            \hline
        \end{tabular}
    \end{center}
    \caption{The input parameters $V_0$ and $d$ used in the CI and DMC calculations of the figure
    \ref{Fig:2}. Also shown are the resulting theoretical values of $\alpha_{RG}$ and $\alpha$.}
    \label{Tab-alpha}
\end{table}
\begin{figure}[t]
\centerline{\includegraphics[clip,width=0.97\columnwidth]{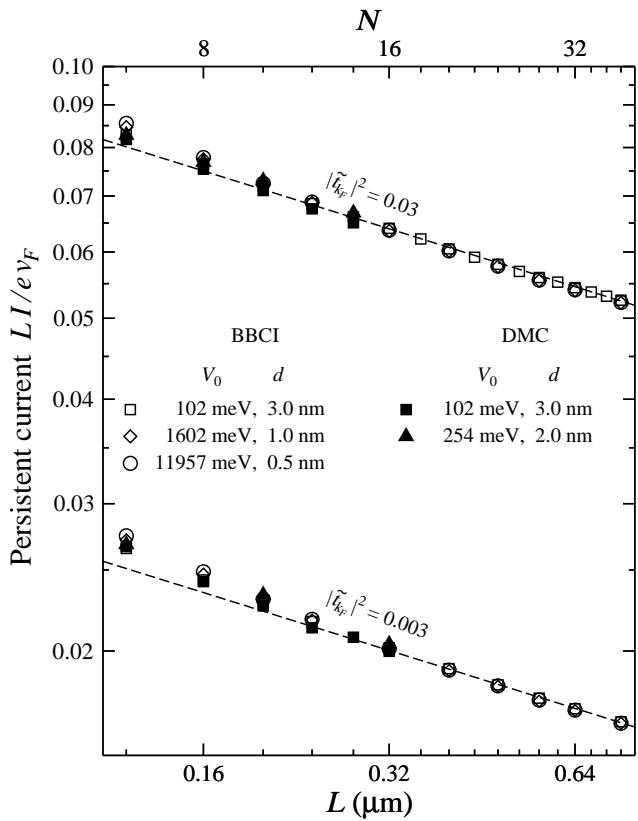}}
\vspace{-0.15cm} \caption{Persistent current $LI/ev_F$ versus the
ring length $L$ for the ring with a single scatterer in the regime
$LI \propto L^{-\alpha}$. Magnetic flux is $\phi = 0.25\phi_0$,
the electron density is $N/L=5 \times 10^7$ m$^{-1}$. We compare
the results for a scatterer with two various transmissions
$|\tilde{t}_{k_{F}}|^2$. We also make comparison for various  e-e
interactions chosen so that the RG formula \eqref{power-Matveev2}
predicts for each considered interaction the same $\alpha_{RG}$.
Specifically, each set $(V_0, d)$ listed in the figure is chosen
so that the RG formula predicts $\alpha_{RG}(V_0,d)=0.2565$. The
dashed lines show the formula $LI \propto L^{-\alpha}$, where
$\alpha = {(1 + 2 \alpha_{RG})}^{1/2} - 1$ and the proportionality
factor is specified in the text. The CI and DMC data are shown by
symbols.} \label{Fig:11}
\end{figure}
\begin{figure}[t]
\centerline{\includegraphics[clip,width=0.97\columnwidth]{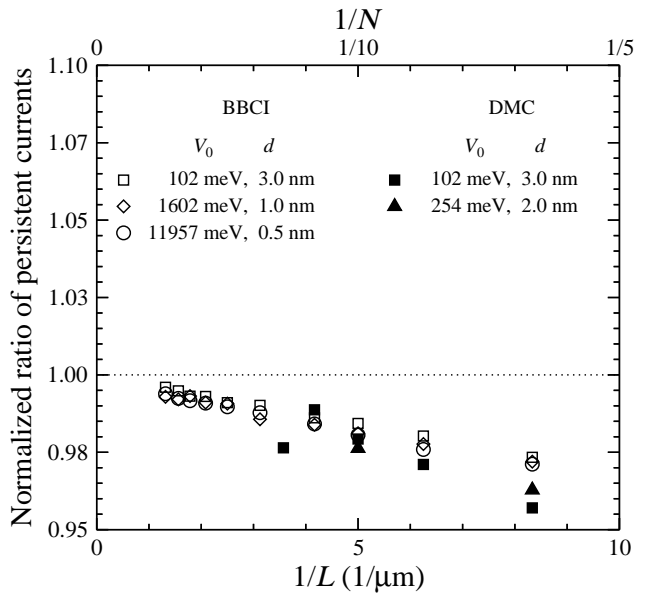}}
\vspace{-0.15cm} \caption{Data from figure \ref{Fig:11} plotted as
the ratio \eqref{I-int-ratio 2}, where $I(\tilde{t}_{k_{F},1})$
and $I(\tilde{t}_{k_{F},2})$ are the persistent currents for
$|\tilde{t}_{k_{F}, 1}|^2 = 0.03$ and $|\tilde{t}_{k_{F}, 2}|^2 =
0.003$, respectively.} \label{Fig:11a}
\end{figure}

In figure \ref{Fig:11} we compare the persistent currents for two
very different transmissions $|\tilde{t}_{k_{F}}|^2$ in order to
demonstrate that the power $\alpha$ is universal - independent on
the choice of $\tilde{t}_{k_{F}}$. Indeed, the CI and DMC data in
figure \ref{Fig:11} exhibit the same $\alpha$ for both
transmissions.

One has to note that the formulae \eqref{power-Matveev} and
\eqref{power-Polyakov} are robust against the choice of the e-e
interaction in the sense, that they give the same $\alpha$ for all
e-e interactions with the same value of $[V(0)-V(2k_F)]/v_F$.
Obviously, this can be the case for many various choices of
$V(x-x')$.

To see whether such robustness exists in our many-body model, in
figure \ref{Fig:11} we also make comparison for various e-e
interactions chosen so that the formula \eqref{power-Matveev2}
gives for various sets $(V_0, d)$ the value
$\alpha_{RG}(V_0,d)=0.2565$. Our CI and DMC data indeed show for
all considered $(V_0, d)$ the decay $LI \propto L^{-\alpha}$,
where $\alpha = {(1 + 2 \alpha_{RG})}^{1/2} - 1$ and
$\alpha_{RG}=0.2565$. In summary, our methods seem to give the
same $\alpha$ for various sets $(V_0, d)$ fulfilling the equation
$\alpha_{RG}(V_0, d)=$\emph{const}. Thus, our methods seem to
confirm the above discussed robustness of the formulae
\eqref{power-Matveev} and \eqref{power-Polyakov} against the
choice of the e-e interaction.

However, we will show later on that this robustness in fact holds
only for the e-e interactions which are very short-ranged [all
sets $(V_0, d)$ in figure \ref{Fig:11} actually belong to this
special limit]. We know that we are in this limit as long as our
methods confirm the formulae \eqref{power-Matveev} and
\eqref{power-Polyakov}. This is still the case in the following
discussion.

We specify the proportionality factor in the formula $LI \propto
L^{-\alpha}$. Following the Introduction, we replace in the
single-particle formula \eqref{I-nonint-approx} the bare amplitude
$\tilde{t}_{k_{F}}$ by the renormalized amplitude $t_{k_{F}}
\propto (\tilde{t}_{k_{F}}/|\tilde{r}_{k_{F}}|) L^{-\alpha}$. We
get
\begin{equation} \label{I-int-heuristic}
I = C \ \frac{ev_F}{2 L} \
\frac{|\tilde{t}_{k_{F}}|}{|\tilde{r}_{k_{F}}|} \ L^{-\alpha}
\sin(2\pi\phi/\phi_0),\quad \alpha>0,
\end{equation}
where $C$ is the proportionality factor. Since $\alpha$ is
universal (independent on $\tilde{t}_{k_{F}}$), the formula
\eqref{I-int-heuristic} implies that
\begin{equation} \label{I-int-ratio}
\frac{I(\tilde{t}_{k_{F},1})}{I(\tilde{t}_{k_{F},2})} =
\frac{|\tilde{t}_{k_{F},1}|/|\tilde{r}_{k_{F},1}|}{|\tilde{t}_{k_{F},2}|/|\tilde{r}_{k_{F},2}|}
\end{equation}
as long as $C$ does not depend on $\tilde{t}_{k_{F}}$. To see that
the CI and DMC data in figure \ref{Fig:11} fulfill the equality
\eqref{I-int-ratio}, we show these data again in figure
\ref{Fig:11a} in terms of the ratio
\begin{equation} \label{I-int-ratio 2}
\frac{|\tilde{t}_{k_{F},2}|/|\tilde{r}_{k_{F},2}|}{|\tilde{t}_{k_{F},1}|/|\tilde{r}_{k_{F},1}|}
\frac{I(\tilde{t}_{k_{F},1})}{I(\tilde{t}_{k_{F},2})},
\end{equation}
which should equal unity for $L \rightarrow \infty$. One sees that
it indeed approaches unity for the CI as well as DMC results. More
precisely, the CI data show a few percent deviation from unity
which tends to disappear with increasing $L$. A small deviation
from unity show also the DMC data where convergence towards unity
is not clear due to the stochastic noise of the Monte Carlo
method.
\begin{figure}[t]
\centerline{\includegraphics[clip,width=\columnwidth]{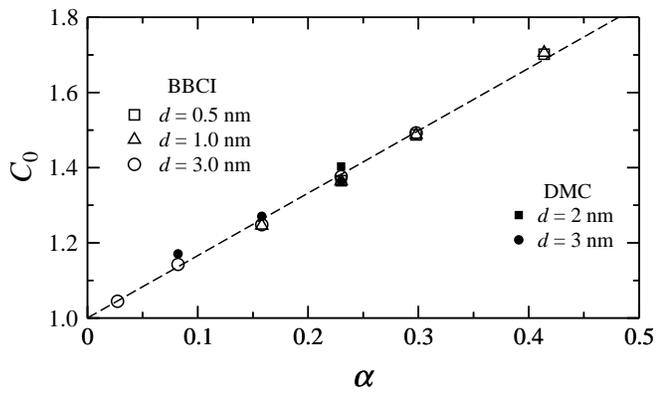}}
\vspace{-0.15cm} \caption{Dependence $C_0(\alpha)$ obtained by
fitting the formula \eqref{I-int-C0} to our CI and DMC data for
$LI/ev_F$. Symbols show the fitted values of $C_0$. The dashed
line is the function $C_0(\alpha) = 1 + 1.66 \alpha$ which fits
the presented data points. } \label{Fig:31}
\end{figure}

Thus, if we ignore a small finite-size effect in figure
\ref{Fig:11a}, our data in figure \ref{Fig:11} are in accord with
equation \eqref{I-int-ratio}. Therefore, we can fit our data in
the $L^{-\alpha}$ regime by the asymptotic formula
\eqref{I-int-heuristic}. For the purpose of fitting it is useful
to define $C \ \equiv {n_e}^{-\alpha}C_0$. Setting this
definition into the formula \eqref{I-int-heuristic} we obtain
\begin{equation} \label{I-int-C0}
\frac{LI}{ev_F} = \frac{1}{2}C_0(\alpha) \
\frac{|\tilde{t}_{k_{F}}|}{|\tilde{r}_{k_{F}}|} \ N^{-\alpha}
\sin(2\pi\phi/\phi_0),
\end{equation}
where $C_0$ depends solely on the parameter $\alpha$ (see below).
Instead of \eqref{I-int-heuristic} we use \eqref{I-int-C0} and we
fit $C_0$ instead of $C$.
\begin{figure}[t]
\centerline{\includegraphics[clip,width=\columnwidth]{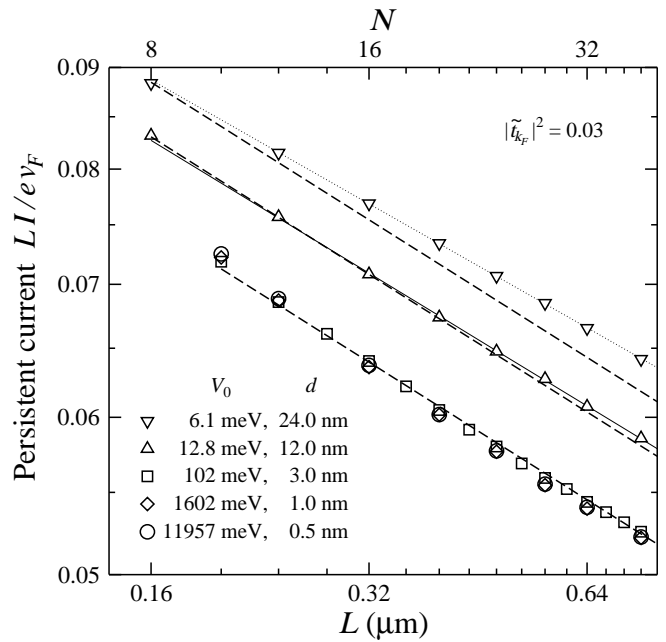}}
\vspace{-0.15cm} \caption{The circles, squares and diamonds are
the BBCI data from the figure \ref{Fig:11}, the triangles and
inverted triangles are the BBCI data for another two sets
$(V_0,d)$. For all considered $(V_0, d)$, the formula
\eqref{power-Matveev2} gives $\alpha_{RG}(V_0,d)=0.2565$ and the
formula \eqref{power-Polyakov} predicts $\alpha = {(1 + 2
\alpha_{RG})}^{1/2} - 1 = 0.23$. However, the dependence $LI
\propto L^{-0.23}$ (the dashed curves with different offset) can
be seen to fit only the BBCI data for $d \leq 3$nm. The BBCI data
for $d = 12$nm are excellently fitted by the curve $LI \propto
L^{-0.222}$ [the full line] and the BBCI data for $d = 24$nm are
fitted by $LI \propto L^{-0.207}$ [the dotted line]. This shows
that the formulae \eqref{power-Matveev} and \eqref{power-Polyakov}
hold only for the e-e interactions which are very short-ranged
(see the text).} \label{Fig:33510}
\end{figure}

The dashed lines in figures \ref{Fig:2} and \ref{Fig:11} show the
dependence \eqref{I-int-C0} fitted to the BBCI data in these
figures. Specifically, $C_0$ is fitted and the function
$N^{-\alpha}$ in \eqref{I-int-C0} is evaluated by using $\alpha =
{(1 + 2 \alpha_{RG})}^{1/2} - 1$, where $\alpha_{RG}(V_0,d)$ is
given by the formula \eqref{power-Matveev2}. The resulting values
of $C_0$ are shown by open symbols in the figure \ref{Fig:31}. The
full symbols in the figure \ref{Fig:31} show the values of $C_0$
obtained when we fit by means of \eqref{I-int-C0} the DMC data in
figures \ref{Fig:2} and \ref{Fig:11}. Also shown are the results
obtained in the same way for another values of $\alpha$. Note that
for all $(V_0, d)$ giving the same $\alpha(V_0, d)$ the obtained
values of $C_0$ are essentially the same. This means that $C_0$
depends solely on the parameter $\alpha$. We also note that $C_0$
does not depend on $\tilde{t}_{k_{F}}$, as has been documented in
the figure \ref{Fig:11a}.


The linear fit in the figure \ref{Fig:31},
\begin{equation} \label{C0alfa}
C_0(\alpha) = 1 + 1.66 \alpha,
\end{equation}
should be viewed as a first estimate of the analytical dependence.
To extract a very precise formula for $C_0(\alpha)$, it would be
desirable to simulate larger systems in order to better suppress
the finite size effect in figure \ref{Fig:11a}.

We conclude that the right-hand side of \eqref{I-int-C0} depends
on a single parameter $\alpha$ via the function $C_0(\alpha)
N^{-\alpha}$, which is single-valued for various $(V_0, d)$ giving
the same $\alpha(V_0, d)$. Our results confirm the decay
$N^{-\alpha}$, where $\alpha$ is given by the formulae
\eqref{power-Polyakov} and \eqref{power-Matveev}.

However, the formulae \eqref{power-Polyakov} and
\eqref{power-Matveev} in fact hold only for very small $d$. In the
figure \ref{Fig:33510} we compare the BBCI data from the figure
\ref{Fig:11} with the BBCI data obtained for another two sets
$(V_0, d)$ with $d$ as large as $12$nm and $24$nm. All considered
$(V_0, d)$ are chosen so that the formulae \eqref{power-Matveev2}
and \eqref{power-Polyakov} give for each $(V_0, d)$ the same value
of $\alpha$, in particular $\alpha = 0.23$. However, it can be
seen, that the dependence $LI \propto L^{-0.23}$ fits only the
BBCI data for $d \leq 3$nm. The BBCI data for $d = 12$nm and $d =
24$nm are excellently fitted by the curves $LI \propto L^{-0.222}$
and $LI \propto L^{-0.207}$, respectively, i.e., with increasing
$d$ our numerically obtained $\alpha$ decreases albeit the value
of $\alpha_{RG}(V_0,d)$ is fixed. This is a strong indication that
the formulae \eqref{power-Polyakov} and \eqref{power-Matveev} hold
only for small $d$. We support this numerical finding by the
following analytical proof.

\begin{table} [t]
    \begin{center}
            \begin{tabular}{|c|c|c||c||c|}
            \hline
            $V_0$ (meV) &   $d$ (nm)    &   $\alpha_{RG}$   &   $\alpha$     &   symbol in Fig. \ref{Fig:22a} \\
            \hline
            \hline
                11      &       3       &       0.0277      &   0.0273       &  $\circ$  \\
                34      &       3       &       0.0855      &   0.0821       &  $\circ$  \\
                68      &       3       &       0.171       &   0.158        &  $\circ$  \\
                1068    &       1       &       0.171       &   0.158        &  $\Box$   \\
                102     &       3       &       0.2565      &   0.230        &  $\circ$  \\
                1602    &       1       &       0.2565      &   0.230        &  $\Box$   \\
                11957   &       0.5     &       0.2565      &   0.230        &$\triangle$\\
                15942   &       1       &       0.342       &   0.298        &  $\Box$   \\
                15942   &       0.5     &       0.342       &   0.298        &$\triangle$\\
                3142    &       1       &       0.5         &   0.414        &  $\Box$   \\
                23445   &       0.5     &       0.5         &   0.414        &$\triangle$\\
            \hline
        \end{tabular}
    \end{center}
    \caption{The input parameters $V_0$ and $d$ used in the BBCI calculations of figure
    \ref{Fig:22a} and the values of $\alpha_{RG}$ and $\alpha$
    resulting from the formulae \eqref{power-Matveev2} and \eqref{power-Polyakov}.
    The last column of the table ascribes a symbol to each set
    $(V_0,d,\alpha_{RG})$.
    These symbols are used in figure \ref{Fig:22a} to show the BBCI data. }
    \label{Tab-alpha2}
\end{table}

\begin{figure}[t]
\centerline{\includegraphics[clip,width=\columnwidth]{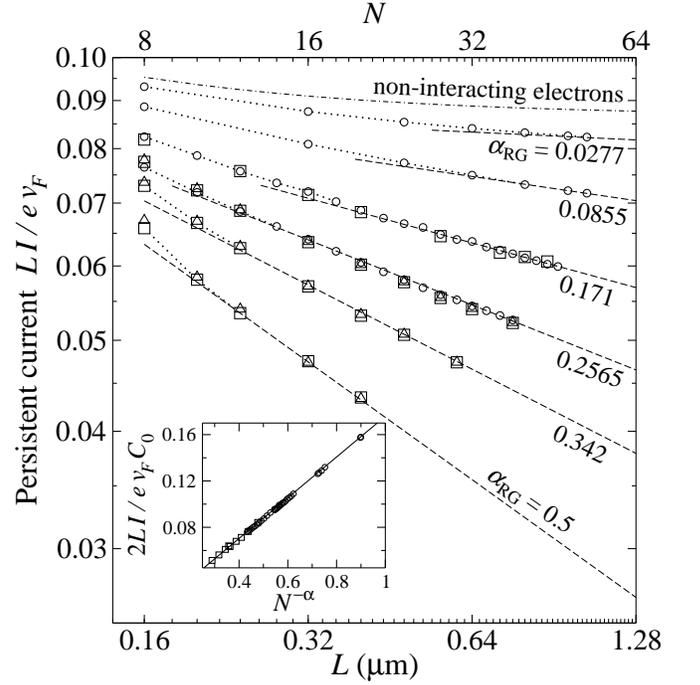}}
\vspace{-0.15cm} \caption{Persistent current versus the ring
length for the same ring as in the figure \ref{Fig:2}, but for a
broader range of the interaction parameters $V_0$ and $d$ (table
\ref{Tab-alpha2}). The circles, squares and triangles show the
results of the BBCI method for various sets $(V_0, d,
\alpha_{RG})$ listed in the table \ref{Tab-alpha2}. The dotted
lines are a guide for eye. The dashed lines show the asymptotic
law $LI \ \propto L^{-\alpha}$ plotted in the form
\eqref{I-int-C0}, where $\alpha = {(1 + 2 \alpha_{RG})}^{1/2} - 1$
and $\alpha_{RG}$ is obtained from \eqref{power-Matveev2}. The
BBCI data in the asymptotic regime are selected, normalized as
$2LI/ev_FC_0$ [with $C_0(\alpha)$ given by the formula
\eqref{C0alfa}], and plotted in inset as a function of the
variable $x \equiv N^{-\alpha}$. The full line is the dependence
$f(x) = \frac{|\tilde{t}_{k_{F}}|}{|\tilde{r}_{k_{F}}|}x$,
predicted by the formula \eqref{I-int-C0}.} \label{Fig:22a}
\end{figure}

In the appendix D we prove analytically that the matrix elements
of the e-e interaction are independent on $d$ for $d \rightarrow
0$ at a fixed value of $\alpha_{RG}(V_0,d)$. In other words, for
$d \rightarrow 0$ the matrix elements depend on a single parameter
$\alpha_{RG}$, otherwise they depend on two parameters  $V_0$ and
$d$. According to the appendix, the limit $d \rightarrow 0$ means
$4k_F^2d^2 \ll 1$. In our calculations $1/2k_F \simeq 3$nm and the
currents in the figure \ref{Fig:33510} are close to each other for
all $d \leq 3$nm. If we increase $\alpha_{RG}$ (for instance as in
the table \ref{Tab-alpha2} and figure \ref{Fig:22a}), we need to
take $d < 1/2k_F$ to see the $d$-independent current convincingly.

Note also another result in figure \ref{Fig:33510}. As $d$ exceeds
$1/2k_F$ the persistent current still decays like $LI \propto
L^{-\alpha}$, but $\alpha$ depends on two parameters $V_0$ and $d$
rather than on a single one, $\alpha_{RG}$. This regime is not
captured by the formulae \eqref{power-Polyakov} and
\eqref{power-Matveev}. We have obtained similar results (not
shown) also for other values of $\alpha_{RG}$.


Consider now a broader range of the e-e interaction parameters
$V_0$ and $d$, listed in the table \ref{Tab-alpha2}. The resulting
persistent currents for these parameters are shown in the figure
\ref{Fig:22a}. All shown BBCI data asymptotically converge to the
Luttinger-liquid behavior described by the formula
\eqref{I-int-C0}, with the power $\alpha$ in accord with the
formulae \eqref{power-Polyakov} and \eqref{power-Matveev}. We wish
to stress the following aspects.

We can write \eqref{I-int-C0} as $\frac{2LI}{ev_FC_0} =
\frac{|\tilde{t}_{k_{F}}|}{|\tilde{r}_{k_{F}}|}x$, where $x \equiv
N^{-\alpha}$ and $C_0(\alpha)$ is given by \eqref{C0alfa}.
Similarly, the BBCI data in the asymptotic regime can be
normalized as $2LI/ev_FC_0$ and plotted as a function of the
variable $x \equiv N^{-\alpha}$. This is done in inset to the
figure \ref{Fig:22a}. Indeed, all BBCI data in inset collapse to a
single curve $f(x) =
\frac{|\tilde{t}_{k_{F}}|}{|\tilde{r}_{k_{F}}|}x$. That a single
linear curve involves the BBCI data for many various $N$, $V_0$
and $d$, is a clear sign of the Luttinger liquid with e-e
interaction depending on one parameter $\alpha_{RG}$. It also
documents that our calculations are reliable for a broad range of
variables $N$, $V_0$ and $d$, the function $C_0(\alpha)$ is
certainly determined with reasonable accuracy.

An interesting finding in figure \ref{Fig:22a} is that for large
$\alpha_{RG}$ the BBCI data show the $LI \propto L^{-\alpha}$
decay already for ten electrons. In other words, the
Luttinger-liquid behavior arises in the system with only ten
particles. For comparison, in the lattice models
\cite{Meden,Enss-2005,Meden-2005,Andregassen} the asymptotic power
law is observed for a much larger number of electrons.

If we compare the formulae \eqref{I-int-C0} and
\eqref{I-nonint-approx}, we see that the interaction modifies the
amplitude $\tilde{t}_{k_{F}}$ as

\begin{equation} \label{transmission-renormalized-CI}
t_{k_{F}} \simeq (\tilde{t}_{k_{F}}/|\tilde{r}_{k_{F}}|)
N^{-\alpha} = (\tilde{t}_{k_{F}}/|\tilde{r}_{k_{F}}|)
{(n_e^{-1}/L)}^{\alpha}.
\end{equation}
This result scales with the length $n_e^{-1}$ while the RG result
\eqref{transmission-Matveev} scales with $d$. What is the origin
of this difference? The RG result \eqref{transmission-Matveev}
should hold for any $d$ obeying the inequality
\eqref{weakinteractionMatveev}, but the inequality
\eqref{weakinteractionMatveev} can in principle be fulfilled also
for $d \lesssim 1/2k_F$, i.e., just for those $d$ for which we
have obtained the formula \eqref{transmission-renormalized-CI}. We
think that the difference between the two results is due to the
different physical models. The formula
\eqref{transmission-renormalized-CI} holds in our continuous
model, with the single-particle energy dispersion being truly
parabolic. The RG formula \eqref{transmission-Matveev} holds in
the model \cite{Matveev}, with the energy dispersion linearized in
the interval $<\varepsilon_F - \hbar v_F/d,\varepsilon_F + \hbar
v_F/d>$ around the Fermi energy. For small $d$ the band width
$\hbar v_F/d$ becomes comparable with $\varepsilon_F$ and the
model \cite{Matveev} has a truly linear dispersion, like the
Luttinger-liquid model \cite{Kane} which also shows scaling by
factor $(d/L)^{\alpha}$. For $d \gg 1/2k_F$ the linearization is
unessential because of $\hbar v_F/d \ll \varepsilon_F$, hence both
models should give the same results. The limit $d \gg 1/2k_F$ is
however not feasible by our numerical methods for computational
reasons.
\begin{figure}[t]
\centerline{\includegraphics[clip,width=\columnwidth]{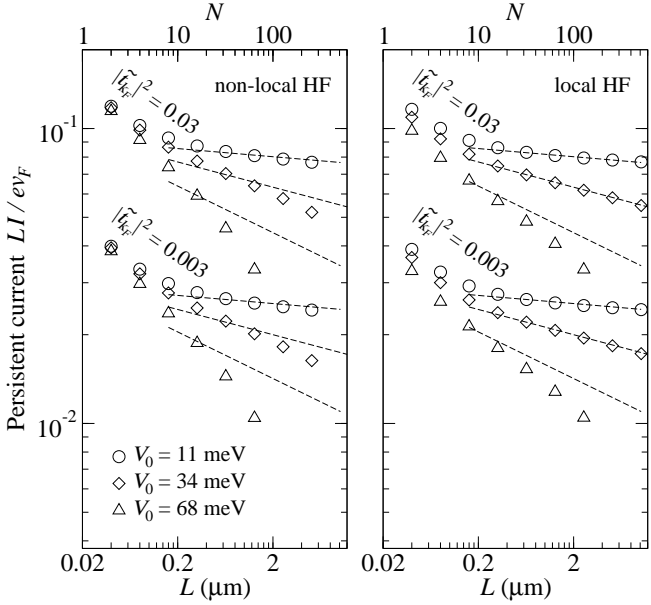}}
\vspace{-0.15cm} \caption{Persistent current $LI/ev_F$ versus ring
length $L$ for the ring with single scatterer, calculated in the
Hartree-Fock approximation for parameters $\phi = 0.25\phi_0$,
$N/L=5 \times 10^7$ m$^{-1}$, $d = 3$nm, and various $V_0$ and
$|\tilde{t}_{k_{F}}|^2$ as listed in the figure. The left panel
shows the nonlocal Hartree-Fock results, the right panel shows the
Hartree-Fock results in the local Fock approximation (see the
text). The Hartree-Fock results are shown by symbols, the dashed
lines show the Luttinger liquid asymptotics $LI \ \propto
L^{-\alpha}$, where $\alpha = {(1 + 2 \alpha_{RG})}^{1/2} - 1$ and
$\alpha_{RG}$ is given by the RG formula \eqref{power-Matveev2}:
$\alpha_{RG}=0.0277, 0.0855$, and $0.1710$ for $V_0 = 11, 34$, and
$68$~meV, respectively.} \label{Fig:3348}
\end{figure}

\subsection{Universal scaling in the Hartree-Fock model}

 Now we discuss the persistent
currents obtained in the self-consistent Hartree-Fock
approximation (Sect.IIC) which ignores correlations except for the
Fock exchange.

In figure \ref{Fig:3348} we show the $I(L)$ dependence, calculated
in the self-consistent Hartree-Fock approximation for the ring
parameters already encountered in our correlated many-body
calculations. The left panel shows the Hartree-Fock data obtained
for the Fock term \eqref{Eqs-Fock} treated as is, the right panel
shows the Hartree-Fock data for the Fock term \eqref{Eqs-Fock}
approximated as \cite{Cohen-98,Cohen-97}
\begin{equation}\label{Eqs-Cohen}
  U_F(x) \simeq - \sum \limits _{n'} \int
  dx'V(x-x'){\rm Re}\left[\psi_{n'}^*(x')\psi_{n'}(x)\right] .
\end{equation}
Unlike the nonlocal interaction \eqref{Eqs-Fock} the approximation
\eqref{Eqs-Cohen} is local - it does not depend on $n$. One
readily obtains \eqref{Eqs-Cohen} from \eqref{Eqs-Fock} by
applying the 'almost closure relation' $\sum_{n'}
\psi_{n'}^*(x')\psi_{n'}(x) \simeq \delta(x-x')$. Once this
approximation is adopted, the final expression for the current has
to be approximated correspondingly \cite{commentI}.

Let us compare our Hartree-Fock calculations with the Luttinger
liquid asymptotics $LI \ \propto L^{-\alpha}$. As can be seen in
the figure \ref{Fig:3348}, both Hartree-Fock approaches show the
$I(L)$ dependence decaying for large $L$ faster than
$L^{-1-\alpha}$. We will see below that for $L \rightarrow \infty$
the decay is in fact exponential. Only for the weak e-e
interaction the Hartree-Fock data show the decay $L^{-1-\alpha}$,
reported also in our previous Hartree-Fock studies
\cite{Nemeth-2005,Gendiar}. However, also in this case we expect
exponential decay at very large $L$.


\begin{figure}[t]
\centerline{\includegraphics[clip,width=\columnwidth]{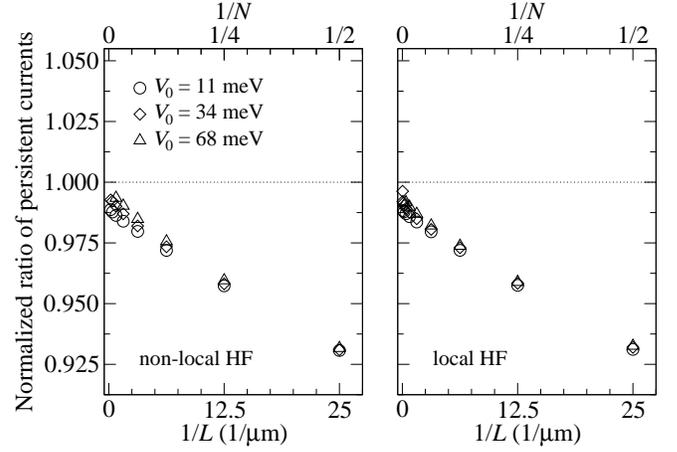}}
\vspace{-0.15cm} \caption{The Hartree-Fock data from the figure
\ref{Fig:3348} plotted as the ratio \eqref{I-int-ratio 2}, where
$I(\tilde{t}_{k_{F},1})$ and $I(\tilde{t}_{k_{F},2})$ are the
persistent currents for the transmission probabilities
$|\tilde{t}_{k_{F}, 1}|^2 = 0.03$ and $|\tilde{t}_{k_{F}, 2}|^2 =
0.003$, respectively.} \label{Fig:3358}
\end{figure}

Now we show that also in the Hartree-Fock approximation the slope
of $\log(I(L))$ becomes for $L \rightarrow \infty$ universal -
dependent only on the e-e interaction. Following equations
\eqref{I-int-heuristic}-\eqref{I-int-ratio 2}, such universality
means that the ratio
$\frac{|\tilde{t}_{k_{F},2}|/|\tilde{r}_{k_{F},2}|}{|\tilde{t}_{k_{F},1}|/|\tilde{r}_{k_{F},1}|}
\frac{I(\tilde{t}_{k_{F},1})}{I(\tilde{t}_{k_{F},2})}$ approaches
unity for $L \rightarrow \infty$. If we redraw in terms of this
ratio the Hartree-Fock data from figure \ref{Fig:3348}, we see (in
the figure \ref{Fig:3358}), that the ratio converges with
increasing $L$ to unity. The stronger the interaction the better
the convergence in the figure \ref{Fig:3358}, for further
improvement a further increase of $L$ is needed. Universality of
the Hartree-Fock results in figure \ref{Fig:3358} is quite similar
to the universality of the correlated results in figure
\ref{Fig:11a}.

\begin{figure}[t]
\centerline{\includegraphics[clip,width=\columnwidth]{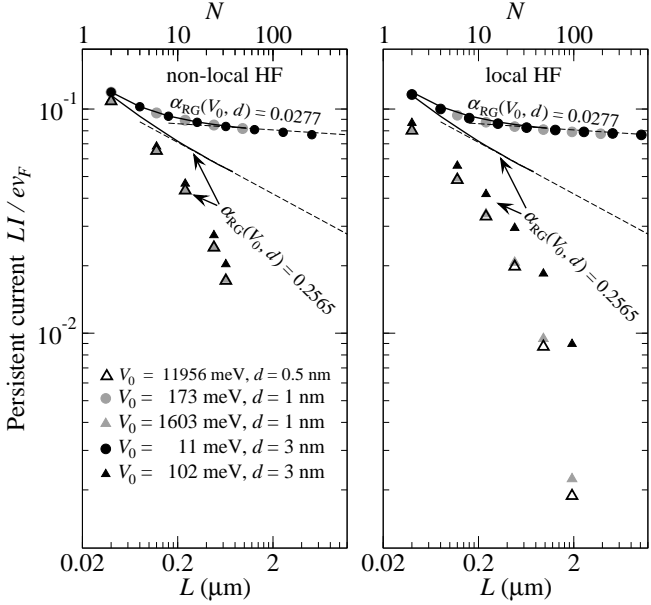}}
\vspace{-0.15cm} \caption{Length-dependence of the persistent
current in the ring with single scatterer, obtained in the
Hartree-Fock approximation. The Hartree-Fock results are shown by
symbols. The ring parameters are: $\phi = 0.25\phi_0$, $N/L=5
\times 10^7$ m$^{-1}$, and $|\tilde{t}_{k_{F}}|^2 = 0.03$. The
weak and strong e-e interactions are simulated for various
$(V_0,d)$ obeying the equations $\alpha_{RG}(V_0,d) = 0.0277$ and
$\alpha_{RG}(V_0,d) = 0.2565$, respectively, where
$\alpha_{RG}(V_0,d)$ is given by the RG formula
\eqref{power-Matveev2}.  Also shown are the corresponding BBCI
data (full lines) and asymptotic $LI \ \propto L^{-\alpha}$ curves
(dashed lines), taken from the figure \ref{Fig:2}.}
\label{Fig:33586}
\end{figure}

Moreover, we would like to show that the Hartree-Fock model
resembles the correlated model also in the following respect. In
the limit $d \rightarrow 0$ the $I(L)$ curve is determined by a
single e-e interaction parameter $\alpha_{RG}$ [i.e., $I(L)$ is
the same for various $(V_0,d)$ giving the same value of
$\alpha_{RG}(V_0,d)$]. This is demonstrated in the figure
\ref{Fig:33586}.

Figure \ref{Fig:33586} shows the Hartree-Fock results for the weak
and strong e-e interactions. Also shown are the corresponding
correlated results from the figure \ref{Fig:2}, in particular the
BBCI data (full lines) and the Luttinger-liquid curves (dashed
lines). Note the following details.
\begin{figure}[t]
\centerline{\includegraphics[clip,width=\columnwidth]{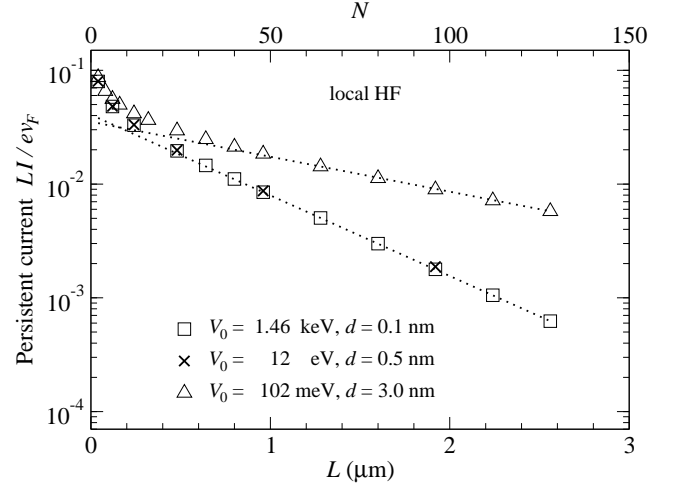}}
\vspace{-0.15cm} \caption{Length-dependence of the persistent
current in the ring with single scatterer, obtained in the local
Hartree-Fock approximation. The Hartree-Fock data are shown by
symbols. The ring parameters are: $\phi = 0.25\phi_0$, $N/L=5
\times 10^7$ m$^{-1}$, and $|\tilde{t}_{k_{F}}|^2 = 0.03$. The
parameters of the e-e interaction, $V_0$ and $d$, are listed in
the figure. The considered sets $(V_0,d)$ obey the equation
$\alpha_{RG}(V_0,d) = 0.2565$, where $\alpha_{RG}(V_0,d)$ is given
by the RG formula \eqref{power-Matveev2}. For large $L$ the
current decays with $L$ linearly, but note that the scale is
semi-logarithmic. The dotted lines fit linearly the Hartree-Fock
data at large $L$.} \label{Fig:33589}
\end{figure}

For weak e-e interaction ($\alpha_{RG} = 0.0277$) the Hartree-Fock
data are robust against various $(V_0,d)$ obeying the equation
$\alpha_{RG}(V_0,d) = 0.0277$. This is illustrated by the
agreement of the Hartree-Fock data for ($V_0=173$meV, $d=1$nm) and
($V_0=11$meV, $d=3$nm). Note also that these Hartree-Fock data
almost reproduce the correlated results at sizes $L$ considered in
the figure. To obtain conclusions valid for any $L$ and any
interaction strength, we have to look at the stronger e-e
interaction.

For strong e-e interaction ($\alpha = 0.2565$) the Hartree-Fock
data fail to agree with the correlated results. However, the
Hartree-Fock data still tend to be robust against various
$(V_0,d$) obeying the equation $\alpha_{RG}(V_0,d) = 0.2565$ in
the limit $d \rightarrow 0$. This universal dependence on a single
e-e interaction parameter $\alpha_{RG}$ is clearly visible both in
the nonlocal and local Hartree-Fock calculation.

Finally, we would like to show that the decay of $I(L)$ in the
Hartree-Fock approximation is exponential for large $L$. This
exponential decay is demonstrated in the figure \ref{Fig:33589},
where the $I(L)$ dependence is presented in the semi-logarithmic
scale. We need to consider large $L$ and strong e-e interaction,
which strongly prolongs the time of the Hartree-Fock computation.
Fortunately, the calculation is feasible in the local
approximation \eqref{Eqs-Cohen}.

The $I(L)$ curves in figure \ref{Fig:33589} do not decay equally
fast, albeit the value of $\alpha_{RG}(V_0,d)$ is the same for all
considered $(V_0,d)$. It can however be seen that if we keep the
same value of $\alpha_{RG}(V_0,d)$ in the limit $d \rightarrow 0$,
then the slope of $I(L)$ is determined solely by the value of
$\alpha_{RG}$. This is again a clear manifestation of the
universal dependence on a single e-e interaction parameter. The
limit $d \rightarrow 0$ is well represented by the data for $d =
0.1$nm.

In summary, our self-consistent Hartree-Fock results show, that
the slope of $\log(I(L))$ in the limit of large $L$ is still
universal - dependent solely on the e-e interaction not on the
strength of the scatterer. Moreover, for $d \rightarrow 0$ this
dependence on the e-e interaction is determined by a single e-e
interaction parameter, specifically by $\alpha_{RG}$ given by the
RG formula \eqref{power-Matveev2}. These features are very similar
to the universal behavior observed in our correlated calculations.
A major difference is that in the Hartree-Fock approximation the
asymptotic decay of $I(L)$ is exponential with $L$, as we have
shown numerically.

\subsection{Reliability and limitations of the CI and DMC}

Both DMC and CI methods in their most rigorous formulations show
exponential scaling of the computer time in the number of
particles. In CI this is due the necessity of infinite size of the
basis and high level of excitations which are required to make the
method size consistent \cite{Szabo,Jensen}. In practice, for not
too large systems, useful and accurate results can be obtained
providing a sufficiently large set of virtual orbitals can be
treated. The question is whether the low-order (doubles, triples,
quadruples) excitations are enough to capture the key many-body
effects. For the DMC method, the exponential
scaling originates in the fermion
sign problem which is in practice avoided by the fixed-node or
fixed-phase approximations. The usefulness of the method crucially
relies on accuracy of the trial functions and ability to
efficiently describe the impact of the many-body effects on
accuracy of the nodes or phase \cite{Foulkes-01}. The DMC practice
shows that very large systems can be treated nevertheless the
fixed-node/phase bias is present and it depends on phenomena of
interest whether it can affect the results.

In subsection IIIA, the largest many-body systems ($48$ electrons)
were simulated by the CI method while only at most $32$ electrons
were simulated by the DMC. This is caused by the fact that the
phase of the Hartree-Fock wave function becomes less accurate with
increasing strength of the interaction and also with increasing
size of the system. This is not too difficult to understand since
in larger and strongly interacting systems the collective
excitations become more complicated and the single determinant
wave function becomes very poor representation of the actual
ground state. The many-body effects description built into the
trial function thus directly determines the accuracy of the
obtained results.


\begin{figure}[t]
\centerline{\includegraphics[clip,width=\columnwidth]{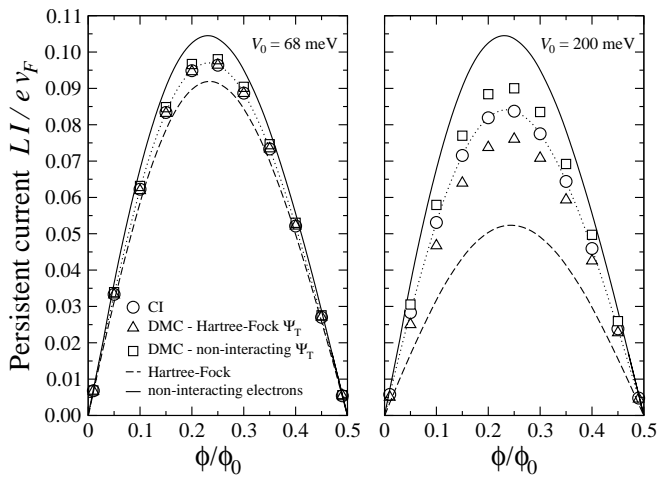}}
\vspace{-0.15cm} \caption{Persistent current versus magnetic flux
for the GaAs ring with $N = 4$ and $L=N/n_e=80$ nm. The
transmission of the scatterer is $|\tilde{t}_{k_{F}}|^2 = 0.03$.
The e-e interaction range is set to $d = 3$ nm and the e-e
interaction magnitude is $V_0 = 68$~meV in the left panel and $V_0
= 200$~meV in the right panel. The triangles show the results of
the fixed-phase DMC calculation with the trial wave function
$\Psi_T(\textbf{X})$ equal to the Slater determinant
\eqref{SlaterHFground} of the self-consistently determined
Hartree-Fock ground state. The squares show the results of the
fixed-phase DMC calculation with the trial wave function
$\Psi_T(\textbf{X})$ equal to the Slater determinant of the
non-interacting ground state. The CI results are shown by circles
connected by the dotted line. For completeness, the results for
the non-interacting ground state are shown in a full line and the
results of the self-consistent Hartree-Fock calculation are shown
in a dashed line.} \label{Fig:3x}
\end{figure}

\begin{figure}[t]
\centerline{\includegraphics[clip,height=12.5cm,width=\columnwidth]{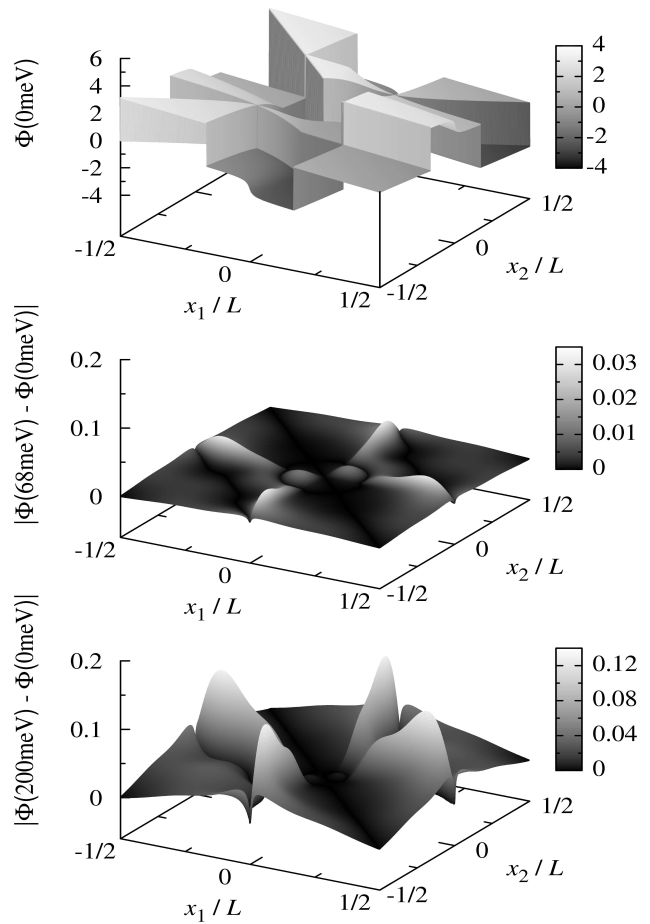}}
\caption{Comparison of various phases from the DMC calculations of
figure \ref{Fig:3x} at $\phi = 0.25\phi_0$. The top panel shows
the phase $\Phi(x_1,x_2,x_3,x_4)$ of the non-interacting trial
wave function $\Psi_T(x_1,x_2,x_3,x_4)$ evaluated at $x_3=-L/6$
and $x_4=L/6$. It is labelled as $\Phi(0 \text{meV})$ since it
holds for $V_0 = 0$meV. The middle panel shows the phase
difference $|\Phi(68 \text{meV})-\Phi(0 \text{meV})|$, where
$\Phi(68 \text{meV})$ is the phase
$\Phi(x_1,x_2,x_3=-L/6,x_4=L/6)$ of the trial wave function
$\Psi_T(x_1,x_2,x_3,x_4)$ obtained  for $V_0 = 68$meV by the
Hartree-Fock method. The bottom panel shows the phase difference
$|\Phi(200 \text{meV})-\Phi(0 \text{meV})|$. We note that the
abrupt change of the phase for $x_i = x_j$ arises because the
Slater determinant changes abruptly its sign for $x_i
\leftrightarrow x_j$.} \label{Fig:o1}
\end{figure}

In figure \ref{Fig:3x} we show the persistent current versus
magnetic flux for the ring with four electrons. The range of the
e-e interaction is $d = 3$nm, the e-e interaction magnitude is
$V_0 = 68$meV in the left panel and $V_0 = 200$meV in the right
panel. The triangles represent the DMC calculation with the trial
wave function $\Psi_T(\textbf{X})$ equal to the Slater determinant
\eqref{SlaterHFground} of the
 Hartree-Fock ground state. The squares show the DMC calculation with the trial wave
function equal to the Slater determinant of the non-interacting
ground state. The results of both DMC calculations are very close
when $V_0 = 68$meV, but they are very different when $V_0 =
200$meV. This suggests that the DMC results for $V_0 = 200$meV are
strongly affected by the phase of the trial wave function. Indeed,
for $V_0 = 68$meV both DMC calculations essentially agree with the
corresponding CI calculation (circles), but for $V_0 = 200$meV the
differences with the CI data are very clear.

It is interesting to see the origin of the fixed-phase biases.
For illustration, in figure \ref{Fig:o1} we compare the phases
involved in the DMC calculations of figure \ref{Fig:3x}. The phase
difference $|\Phi(68 \text{meV})-\Phi(0 \text{meV})|$ is small and
this is why the corresponding DMC currents in the left panel of
figure \ref{Fig:3x} almost coincide. The phase difference
$|\Phi(200 \text{meV})-\Phi(0 \text{meV})|$ is large and the
corresponding DMC currents in the right panel of figure
\ref{Fig:3x} clearly differ.

\begin{figure}[t]
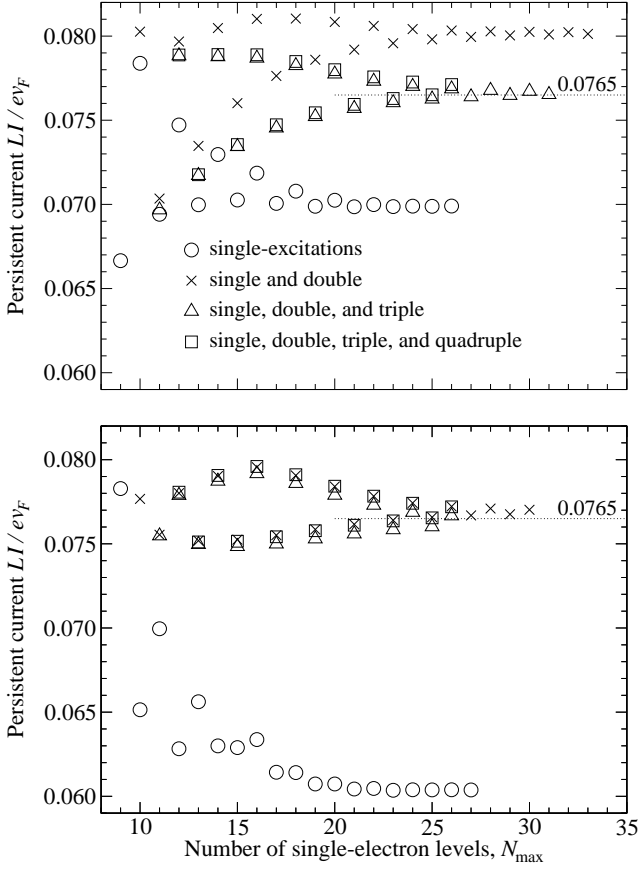

\centerline{\includegraphics[clip,width=\columnwidth]{PRB_2008_Fig14a.eps}}
\centerline{\includegraphics[clip,width=\columnwidth]{PRB_2008_Fig14b.eps}}
\caption{Convergence of the persistent current in the FCI
calculation in dependence on the number of the considered
single-electron levels for the single-excitations, single- and
double-excitations, etc. The top panel shows the currents obtained
from the formula $I = \left< \Psi_0 \right| \hat{I} \left| \Psi_0
\right>$, the bottom panel shows the currents obtained in the same
FCI calculation from the formula $I = -\partial E_0 /
\partial \phi$. The results are shown
by symbols. They were obtained for the ring with parameters $N =
8$ and $L=N/n_e=0.16$ $\mu$m, penetrated by magnetic flux $\phi =
0.25\phi_0$. The transmission of the scatterer is
$|\tilde{t}_{k_{F}}|^2 = 0.03$. The e-e interaction is given by
$V_0 = 102$~meV and $d = 3$ nm.} \label{Fig:konvergencia_FCIprud}
\end{figure}

\begin{figure}[t]
\centerline{\includegraphics[clip,width=\columnwidth]{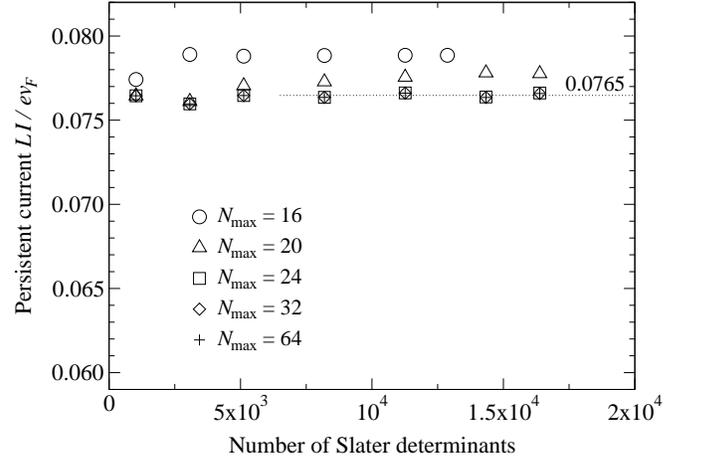}}
\vspace{-0.15cm} \caption{Convergence of the persistent current in
the BBCI calculation for various $N_{max}$ in dependence on the
number of the Slater determinants involved in the expansion
\eqref{SLaterexpansion}. We recall that the determinants are
selected by using the bucket-brigade algorithm of Appendix C and
$N_{max}$ is the number of the considered single-electron levels.
In this BBCI calculation the current is evaluated by using the
formula $I = \left< \Psi_0 \right| \hat{I} \left| \Psi_0 \right>$
and the parameters of the ring are the same as in the FCI
calculation of the preceding figure. The value $0.0765$ is the
average from the (well saturated) data for $N_{max} \ge 24$.}
\label{Fig:333}
\end{figure}

We conclude that the main limitation of our DMC results is the
Hartree-Fock approximation of the correct phase, which
deteriorates with increase of the e-e interaction and system size.
In particular, we see in the figure \ref{Fig:3x} that the DMC with
the Hartree-Fock trial wave function underestimates the persistent
current. For instance, in figure \ref{Fig:2} the DMC data for the
$\alpha_{RG} = 0.2565$ exhibit a weak underestimation which is of
the same origin.

Reliability of the CI results depends on how the results converge
with increasing the number of the Slater determinants in the
expansion \eqref{SLaterexpansion}.

The figure \ref{Fig:konvergencia_FCIprud} shows a typical
convergence process in the FCI calculation. We recall that we add
into the expansion \eqref{SLaterexpansion} first the Slater
determinant of the ground state, then all determinants with a
single excited electron (the single-excitations), all determinants
with two excited electrons (the double-excitations), etc. The
figure \ref{Fig:konvergencia_FCIprud} demonstrates how the
persistent current saturates with increasing the size of the
single-electron basis (the number of the considered
single-electron levels) for the single-excitations only, for the
single- and double-excitations, etc. Moreover, the top and bottom
panel compare convergence of the persistent currents calculated by
means of the formulae $I = \left< \Psi_0 \right| \hat{I} \left|
\Psi_0 \right>$ and $I = -\partial E_0 /
\partial \phi$, respectively. It can be seen that both approaches converge to the same
final result when the single-, double-, triple-, and
quadruple-excitations are considered. However, precision of the
results in the bottom panel is good already for the single- and
double-excitations while in the top panel also the
triple-excitations are needed to achieve a comparable precision.

The FCI calculations with the formula $I = -\partial E_0 /
\partial \phi$ therefore consume much less computer time and memory. The FCI
data presented in this text were obtained by means of the formula
$I = -\partial E_0 /
\partial \phi$ and by considering the single- and double-excitations.
In a few cases, sufficiency of the achieved convergence was
confirmed by adding the triple-excitations or even the
quadruple-excitations, with a similar success as in the figure
\ref{Fig:konvergencia_FCIprud}.

The figure \ref{Fig:333} shows typical convergence of the BBCI
calculation in dependence on the number of the Slater determinants
involved in the expansion \eqref{SLaterexpansion}. The persistent
currents in the BBCI converge to the value $0.0765$, in accord
with the value reached by the FCI data in figure
\ref{Fig:konvergencia_FCIprud}. The BBCI data in this text were
obtained by using the formula $I = \left< \Psi_0 \right| \hat{I}
\left| \Psi_0 \right>$. The BBCI relying on the formula $I =
-\partial E_0 /
\partial \phi$ gives the same results (not shown), but the computational time is about twice
longer.

\begin{figure}[t]
\centerline{\includegraphics[clip,width=\columnwidth]{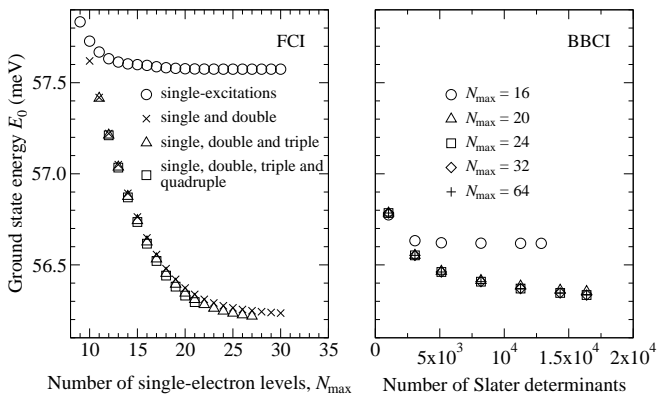}}
\vspace{-0.15cm} \caption{The left panel shows how the
ground-state energy $E_0$ converges in the FCI calculation of
figure \ref{Fig:konvergencia_FCIprud}. The right panel shows how
$E_0$ converges in the BBCI calculation of figure \ref{Fig:333}.}
\label{Fig:3347}
\end{figure}

In figure \ref{Fig:3347} we demonstrate convergence of the
ground-state energy $E_0$ in the FCI (left panel) and BBCI
calculation (right panel). In both cases $E_0$ converges to a
certain minimum. Here $E_0$ is better minimized in the FCI
calculation than in the BBCI, but we note that the BBCI result
would eventually converge (for a significantly larger number of
determinants) to the same minimum. In both cases, however, we need
to add more determinants to achieve a perfectly saturated $E_0$.
In contrast to this, if we look at the convergence of the
corresponding persistent currents (figures
\ref{Fig:konvergencia_FCIprud} and \ref{Fig:333}, respectively),
we see that it is satisfactory finished both in the FCI and BBCI.
Thus, as long as we are interested in the persistent current, it
is economical to look at the convergence of the current rather
than at the convergence of $E_0$.

\begin{figure}[t]
\centerline{\includegraphics[clip,width=\columnwidth]{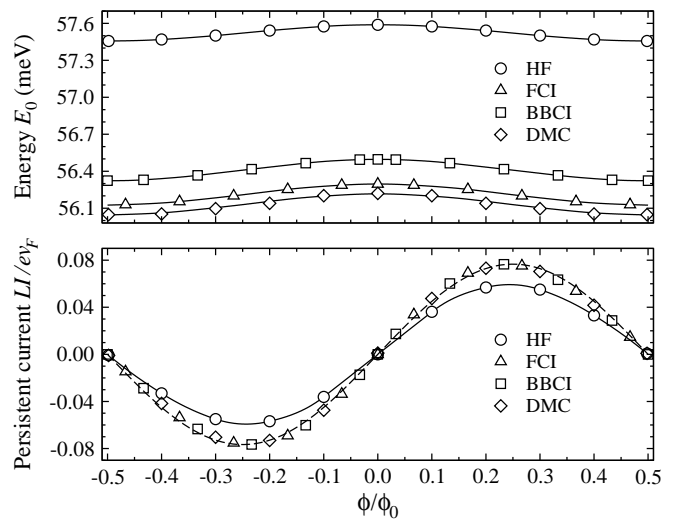}}
\vspace{-0.15cm} \caption{The ground-state energy $E_0$ and
persistent current as functions of magnetic flux for the ring with
parameters $N = 8$ and $L=N/n_e=0.16$ $\mu$m. The transmission of
the scatterer is $|\tilde{t}_{k_{F}}|^2 = 0.03$. The parameters of
the e-e interaction are $V_0 = 102$~meV and $d = 3$ nm, they give
$\alpha_{RG} = 0.2565$. The CI, DMC and Hartree-Fock (HF) results
are shown by symbols. The full lines are a guide for eye, the
dashed line shows the formula \eqref{I-int-C0}, with $\alpha = {(1
+ 2 \alpha_{RG})}^{1/2} - 1$ in the function $N^{-\alpha}$ and
with the value of $C_0$ taken from  figure \ref{Fig:31}.}
\label{Fig:3}
\end{figure}

In figure \ref{Fig:3} we show the ground-state energy and
persistent current as functions of magnetic flux for the same
parameters as in figures \ref{Fig:konvergencia_FCIprud},
\ref{Fig:333}, and \ref{Fig:3347}. The CI and DMC results for the
currents are in good agreement. However, the CI and DMC
ground-state energies exhibit small differences which deserve a
comment. In particular, in spite of the fixed-phase approximation
the DMC ground-state energy shows the best minimum. The reasons
for this are the following. First, the FCI results in the figure
\ref{Fig:3} were obtained by including the single, double, and
triple-excitations (see the discussion of figures
\ref{Fig:konvergencia_FCIprud} and \ref{Fig:3347}). The
quadruple-excitations, etc., would shift the FCI ground-state
energy to a slightly lower value. Second, the presented BBCI
results were obtained for $N_{max} = 64$ and for $1.6 \times 10^4$
Slater determinants (see figures \ref{Fig:333} and
\ref{Fig:3347}). Inclusion of more determinants would shift the
BBCI ground-state energy to a lower value.

It is tempting to think that the result showing the lowest
ground-state energy is the best one. Such criterion indeed follows
from the variational principle if we compare various methods
applied to the same Hamiltonian. Here the CI methods are applied
to the exact Hamiltonian but the DMC is applied to the effective
Hamiltonian. If the ground-state energy of the effective
Hamiltonian is lower than the ground-state energy of the exact
Hamiltonian, this is not in conflict with the variational
principle. It should also be stressed that the convergence of
other properties such as expectation value of the current operator
is not necessarily governed by the same behavior as the energy
convergence. We believe that our CI results are closer to the true
ones due to the good CI convergence demonstrated above. On the
other hand the crude Hartree-Fock wave functions used in the
fixed-phase approximation are probably not accurate enough to
provide accurate currents for strongly interacting limit. This
remains an interesting point for future studies since more
accurate trial wave functions can be constructed by means of
pairing orbitals and pfaffians, expansions in pfaffians and/or
using backflow (dressed) many-body coordinates \cite{Bajdich08}.

\section{Summary and concluding remarks}

We have studied the persistent current of the interacting spinless
electrons in a continuous 1D ring with a single strongly
reflecting scatterer. We have included correlations by solving the
Schrodinger equation for several tens of electrons interacting via
the pair interaction $V(x - x') = V_0 \, \exp(- \left| x - x'
\right|/d)$. Our aim was to solve this continuous many-body
problem without any physical approximation and to examine
microscopically the power-law behavior of the Luttinger liquid. We
have used advanced CI and DMC methods which, unlike the
Luttinger-liquid model, do not rely on the Bozonization technique.

In the past, similar studies were performed by numerical RG
methods in the lattice model
\cite{Meden,Enss-2005,Meden-2005,Andregassen}. Our methods do not
rely on the RG techniques and thus serve as independent
confirmation of such approaches. Moreover, dealing with the
continuous model, we can vary the range of the e-e interaction in
order to test the robustness of the Luttinger-liquid power laws
against various shapes of $V(x - x')$. This point was not
addressed in the lattice-model studies as the range of the
interaction was usually fixed to the on-site interaction and/or to
the nearest-neighbor-site interaction. Our major findings are:

(i) Our CI and DMC calculations confirm that the persistent
current exhibits the asymptotic power-law behavior of the
Luttinger liquid, $I \propto L^{-1-\alpha}$.

(ii) In our continuous model, the question whether the power
$\alpha$ is determined by a single e-e interaction parameter
$\alpha_{RG}$ is addressed by using various shapes of $V(x-x')$
giving the same value of $\alpha_{RG}$. Our numerical values of
$\alpha$ confirm the theoretical formula $\alpha = {(1 + 2
\alpha_{RG})}^{1/2} - 1$ with $\alpha_{RG}$ given by the RG
expression \eqref{power-Matveev}, but only if the range of $V(x -
x')$ is small ($d \lesssim 1/2k_F$).

(iii) The CI data for $\alpha_{RG} \gtrsim 0.3$ show onset of the
asymptotics $I \propto L^{-1-\alpha}$ already for ten electrons.
In other words, the Luttinger-liquid behavior emerges in the
system with only ten particles. For comparison, a far much larger
number of electrons is needed to observe the asymptotic power law
in the lattice model
\cite{Meden,Enss-2005,Meden-2005,Andregassen}. To understand
origin of this difference, it would be desirable to study smooth
transition from the lattice model to the continuous model.


(iv) We have treated the e-e interaction in the self-consistent
Hartree-Fock approximation. We observe for large $L$ the
exponentially decaying $I(L)$ instead of the power law. However,
the slope of $\log(I(L))$ still depends solely on the parameter
$\alpha_{RG}$ given by the RG formula \eqref{power-Matveev}, as
long as the range of $V(x - x')$ approaches zero.

(v) We have discussed the reliability and limits of our CI and DMC
calculations. The CI methods appear to provide convergent results
for the sizes and strengths of interactions we have studied. The
reliability of the FCI and BBCI results is easy to analyze in this
case and both methods converge to the same results although the
BBCI can treat larger systems. The main limitation of our DMC
calculations is the accuracy of Hartree-Fock trial function phase
which has been employed in the fixed-phase calculations and
appears to be a rather poor approximation for strong e-e
interaction.

\begin{figure}[t]
\begin{center}
  \includegraphics[clip,width=\columnwidth]{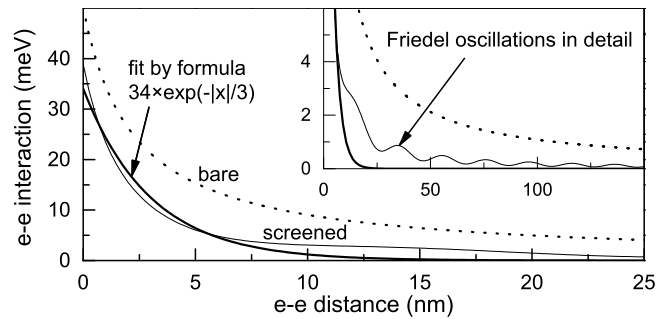}
\end{center}
\caption{Electron-electron interaction versus distance in the GaAs
1D system. The dotted line shows the bare e-e interaction
$V_{\text{bare}}(x - x') = \frac{e^2}{4\pi \epsilon} \, \frac{1}
{\left| x-x' \right| + x_0}$ for $\epsilon = 12.5\epsilon_0$ and
$x_0 = 3$ nm, where the cutoff $x_0$ mimics the finite wire
thickness. The thin full line shows the potential of a single
electron screened by the free 1D electron gas, calculated in the
Hartree picture \cite{comment2}. The full line is the best fit by
formula $V(x - x') = V_0 \,  e^{- \left| x - x' \right|/d}$, where
$d = 3$nm essentially coincides with the value of $1/2k_F$. We
expect that external gates would further increase the screening,
i.e., it is indeed meaningful to focus on the range $d \lesssim
1/2k_F$ as we did in our work. The Friedel oscillations in the
figure are artefact of static screening and would be further
suppressed by the gates. } \label{Fig:Pot1}
\end{figure}

Although our work is not aimed to address experimental aspects,
nevertheless, one of our findings might be a motivation for
experimental work. It might be interesting to observe onset of the
Luttinger-liquid behavior in the 1D system with a small number of
strongly interacting electrons, e.g. with only ten electrons as
predicts our work. In this respect we mention, that the screened
e-e interaction \eqref{VeeExp} with the parameters $V_0$ and $d$
considered in our calculations is quite realistic. This is
documented in figure \ref{Fig:Pot1}, where the model interaction
\eqref{VeeExp} is compared with the microscopic Hartree screening
of the bare e-e interaction.


\section{Acknoledgement}
The IEE group was supported by the grant APVV-51-003505, grant
VEGA 2/6101/27, and ESF project VCITE. L. M. thanks for support
from NSF and DOE.

\section*{Appendix A: Numerical algorithm for calculation of amplitudes $t_k$ and $r_k$}

Consider again the segment $\langle-L/2,L/2\rangle$ embedded
between two perfect semi-infinite leads. Outside the segment the
electron wave function reads

\begin{equation} \label{fromtheleft}
\varphi_k(x)=e^{ikx}+r_k e^{-ikx}, \ \varphi_k(x)=t_k e^{ikx}
\end{equation}
for the electron impinging the segment from the left and

\begin{equation} \label{fromtheright}
\varphi_{-k}(x)=t'_k e^{-ikx}, \ \varphi_{-k}(x)=e^{-ikx}+r'_k
e^{ikx}
\end{equation}
for the electron impinging the segment from the right, where $k
>0$, $r_k$ is the reflection amplitude, and $t_k$ is the
transmission amplitude. Therefore, the boundary conditions for
elastic tunneling through the segment $\langle-L/2,L/2\rangle$
have a standard form

\begin{equation} \label{BCondgt}
\varphi_k(-\frac{L}{2})=e^{-ik\frac{L}{2}}+r_k e^{ik\frac{L}{2}},
\ \varphi_k(\frac{L}{2})=t_k e^{ik\frac{L}{2}},
\end{equation}
\begin{equation} \label{BCondls}
\varphi_{-k}(- \frac{L}{2})=t'_k e^{ik\frac{L}{2}}, \
\varphi_{-k}(\frac{L}{2})=e^{-ik\frac{L}{2}}+r'_k
e^{ik\frac{L}{2}} .
\end{equation}
To calculate $t_k$ and $r_k$, we have to solve equation
\eqref{independent Eqs-Schroding-2} as a tunneling problem with
boundary conditions \eqref{BCondgt}. We write equation
\eqref{independent Eqs-Schroding-2} in the discrete form \eqref
{Stoermer}. Consider now the electron with $k>0$. It leaves the
segment $\langle-L/2,L/2\rangle$ at $x = L/2$ as a free wave $
\propto e^{ikL/2}$. We can thus initialize the scheme \eqref
{Stoermer} at $x = L/2 + \Delta$ and $x = L/2$ by using

\begin{equation} \label{Init} \nonumber
\varphi_k ^{\text{num}} \left( L/2+ \Delta \right) = e^{ik(L/2+
\Delta)} , \quad \varphi_k ^{\text{num}} \left( L/2 \right) =
e^{ikL/2} .
\end{equation}
By means of \eqref {Stoermer} we generate the numerical solution
$\varphi_k ^{\text{num}} \left( x_j \right)$ starting at $x = L/2
- \Delta$ and ending at $x = -L/2$. Since $\varphi_k \left( L/2
\right) = t_k \varphi_k ^{\text{num}} \left( L/2 \right)$, the
correctly scaled result is $\varphi_k \left(x \right) = t_k
\varphi_k ^{\text{num}} \left(x \right)$, where $t_k$ is yet
unknown. We can readily express $t_k$ and $r_k$ from the boundary
condition

\begin{equation} \label{BCNum1}
t_k \, \varphi_k ^{\text{num}} \left( x \right) = e^{ikx} + r_k \,
e^{-ikx} , \quad  x=-\frac{L}{2},
\end{equation}
and from the flux continuity equation

\begin{equation} \label{BCNum2}
t_k \frac {d} {dx}\, \varphi_k ^{\text{num}} \left( x \right) =
\frac {d} {dx}\, \left[
  e^{ikx} + r_k \, e^{-ikx}
\right] , \quad  x=-\frac{L}{2}.
\end{equation}
A similar procedure can be used to obtain the solution
$\varphi_{-k} \left(x \right)$ together with the amplitudes $t'_k$
and $r'_k$.

\section*{Appendix B: Iteration steps of the Hartree-Fock calculation}

For numerical purposes it is useful \cite{comment1} to replace the
potential $\gamma\delta(x)+U_H(x)+U_F(n,x)$ in the Hartree-Fock
equation \eqref{Eqs-Schroding-2} by the potential

\begin{multline}\label{Eqs-Potencial-iter}
 U_{new}(n,x) = f[\gamma\delta(x)+U_H(x)+U_F(n,x)]
\\
+(1-f)U_{old}(n,x) \,.
\end{multline}
In a given iteration step the equation \eqref{Eqs-Schroding-2} is
solved with the potential $U_{new}(n,x)$, where $U_{old}$ is
$U_{new}$ from the previous iteration step, and $f<1$ is a
properly chosen weight. In the first iteration step we set
$U_{new} \equiv \gamma\delta(x)$.

\begin{figure}[!htb]
  \begin{center}
    \includegraphics[clip,width=7.8cm]{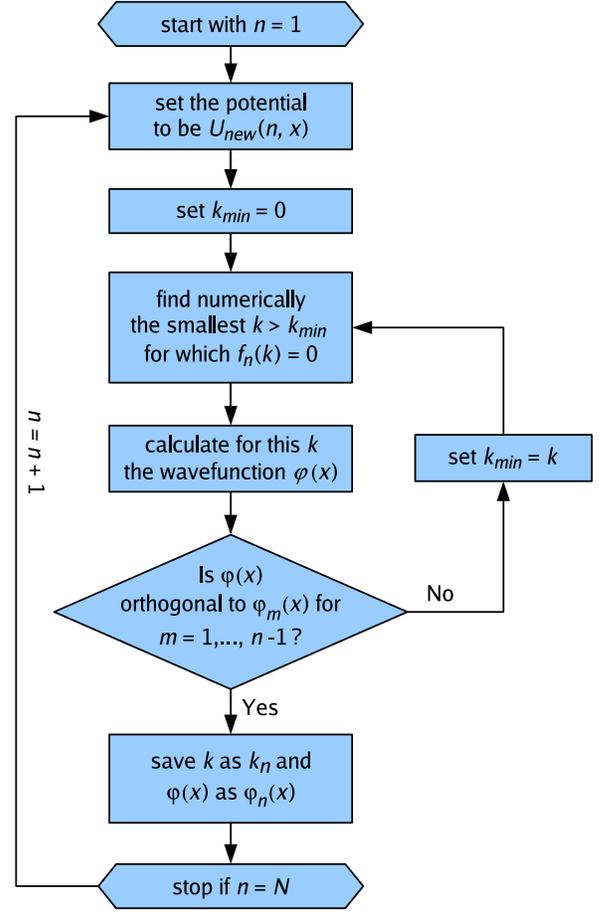}
  \end{center}
  \caption{Flowchart of a single Hartree-Fock iteration step as described in the text.}\label{Fig-FlowChart}
\end{figure}

The iteration step works as follows. We search for the
Hartree-Fock energies $\varepsilon_n$ by solving the equation
\eqref{Eqs-trancedent}. Precisely, we solve the equation $f_n(k) =
0$, where

\begin{equation}\label{Eqs-Transcendent-c-1}
  f_n(k)={\rm Re}\{e^{-ikL}/t_k\}-\cos(2\pi\phi/\phi_0)
\end{equation}
and $t_k$ is the transmission evaluated for the potential
$U_{new}(n,x)$. Clearly, the equation $f_n(k) = 0$ is fulfilled
for many different values of the variable $k$. Among these values
we choose as physically correct only those $k = k_n$ which fulfill
simultaneously two conditions. The first condition is the
inequality $\varepsilon_1<\varepsilon_2<\dots<\varepsilon_N$,
where $\varepsilon_1, \varepsilon_2, \dots \varepsilon_N$ are the
energies of the $N$ lowest levels, each of them given as
$\varepsilon_n=\hbar^2k_n^2/2m$. The second condition is that the
wave functions $\varphi_n(x)$ of these $N$ lowest levels are
mutually orthogonal. The flowchart of the iteration step is shown
in the figure \ref{Fig-FlowChart}.

After finishing the iteration step we set the obtained
$\varphi_n(x)$ into the potentials $U_H(x)$ and $U_F(n,x)$ and we
obtain $U_{new}(n,x)$ for the following iteration step. We repeat
the iteration steps until the energies $\varepsilon_n$ do not
change anymore.

We note that we calculate the transmission $t_k$ for the given
$U_{new}(n,x)$ by means of the algorithm from Appendix A. The wave
function $\varphi_n(x)$ of the energy level
$\varepsilon_n=\hbar^2k_n^2/2m$ is calculated by using the
procedure described in the last paragraph of Section II.A.

\section*{Appendix C: Implementation of BBCI}

Here we outline the BBCI algorithm \cite{Indlekofer}. As mentioned
in section II.E, we determine the single-particle basis
$\psi_j(x)$ with the ladder of the single-particle-energy levels
$\varepsilon_j$ and we consider only the first $N_{max}$ levels by
introducing the upper energy cutoff as illustrated in the figure
\ref{Fig:CI}. Due to the cutoff, the expansion
\eqref{SLaterexpansion} contains the finite number of the Slater
determinants $\chi_n$. This number, equal to
$\left({N_{max}}\atop{N}\right)$ by simple combinatorics, is
usually huge. The question is how to asses the importance of all
$\left({N_{max}}\atop{N}\right)$ determinants and to omit those of
them which are unimportant. In principle, the importance criterion
adopted in section II.E requires to compute the energy $\left<
\chi_n \right| \hat{H} \left| \chi_n \right>$ for all
$\left({N_{max}}\atop{N}\right)$ determinants and to order the
obtained numerical values $\left< \chi_n \right| \hat{H} \left|
\chi_n \right>$ increasingly. Since
$\left({N_{max}}\atop{N}\right)$ is huge, such direct approach
still consumes enormous amount of the computer memory and it is
also time-consuming to order increasingly a series of
$\left({N_{max}}\atop{N}\right)$ numbers. The bucket-brigade
algorithm \cite{Indlekofer} eliminates these difficulties as
follows.

\begin{figure}[t]
  \begin{center}
    \includegraphics[clip,width=7.8cm]{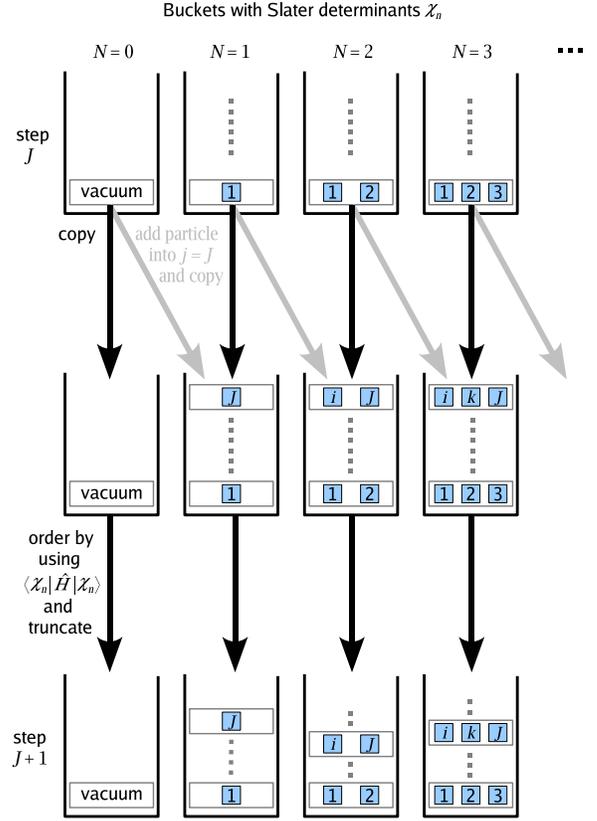}
  \end{center}
  \caption{Visualization of the recursion step $J\rightarrow J+1$ discussed in the text:
  Copy the remaining (after truncation) Slater determinants from the $N$-particle bucket in step $J$
  into the $N$-particle bucket in step $J+1$. Copy the remaining (after truncation)
  Slater determinants from the $(N-1)$ -particle bucket in step $J$
  into the $N$-particle bucket in step $J+1$, but, when copying, enlarge each Slater determinant
  by creating a particle in the state $j = J$. Truncate
  with the help of the measure $\left< \chi_n \right| \hat{H} \left| \chi_n \right>$
  the Slater determinants of each bucket in the step $J+1$. Go to the step $J+2$. Stop
  the procedure after $J$ reaches $N_{max}$. The squares represent the occupied single-particle
  states $j=1,2,\dots,J$. The number of the particles in the bucket is labelled by $N$.}
  \label{Obr-BBCI-scheme}
\end{figure}

We use the single-particle states $\psi_j(x)$ to generate the
Slater determinants $\chi_n$ via a recursion algorithm (Fig.
\ref{Obr-BBCI-scheme}) which is based on the sequence of buckets
$S_{J,N}$ with $J=0,\cdots,N_{max}$ and $N=0,\cdots,N_{max}$,
where $N$ is the number of the particles in the bucket and $J$
labels the recursion step $J\rightarrow J+1$. Each bucket
$S_{J,N}$ contains only the Slater determinants of $N$ particles
which occupy the single-particle states $\psi_j(x)$ with
$j=0,\cdots,J-1$. In each recursion step $J\rightarrow J+1$, we
first expand the old buckets $S_{J,N}$ by adding the Slater
determinants from the buckets $S_{J,N-1}$, but with an extra
particle created in the single-particle state $j=J$. After this
expansion, the buckets $S_{J+1,N}$ are truncated by selecting only
the most important Slater determinants in accord with our
importance criterion based on the chosen measure of importance,
which is chosen as $\left< \chi_n \right|\hat{H}\left| \chi_n
\right>$ in this paper. Owing to this algorithm our importance
criterion is applied only to the determinants in the bucket rather
than to all $\left({N_{max}}\atop{N}\right)$ determinants.

\section*{Appendix D: Matrix elements of electron-electron interaction}

If we expand the function $f(x+y)$ into the Taylor series around
$x$, we can write

\begin{equation}\label{Eqs-interaction-integral}
    \int\limits_{-L/2}^{L/2} {\rm d}y V(y)f(x+y)=\sum_{n=0}^\infty v_nf^{(n)}(x)\,,
\end{equation}
where $f^{(n)}(x) = \frac{\partial^n}{\partial x^n} f(x)$ and

\begin{equation}\label{Eqs-ee-transform}
    v_n=\frac{1}{n!}\int\limits_{-L/2}^{L/2} {\rm d}y V(y)y^n\,.
\end{equation}
Setting the short range interaction $V(y)=V_0e^{-|y|/d}$ into
\eqref{Eqs-ee-transform} we get

\begin{equation}
    v_n=\frac{V_0}{n!}\int\limits_{-L/2}^{L/2} {\rm d}y e^{-|y|/d}y^n=\frac{V_0}{n!}
        [1+(-1)^n]\int\limits_{0}^{L/2} {\rm d}y e^{-y/d}y^n
\end{equation}
and after integration per partes we obtain

\begin{equation}
    v_n=V_0d\,[1+(-1)^n]\!\left[d^n-e^{-L/2d}\sum_{m=0}^n\frac{d^m(L/2)^{n-m}}{(n-m)!}\right]\,.
\end{equation}
Setting this $v_n$ back into \eqref{Eqs-interaction-integral},
reordering the summations over $m$ and $n$, identifying the
particular Taylor series of $f^{(n)}(x\pm L/2)$, and using the
periodicity condition $f^{(n)}(x\pm L)=f^{(n)}(x)$, we obtain the
relation

\begin{multline}
    \int\limits_{-L/2}^{L/2} {\rm d}y V_0e^{-|y|/d}f(x+y)=\\
    2V_0d\sum_{n=0}^\infty d^{2n}\left[f^{(2n)}(x)-e^{-L/2d}f^{(2n)}(x-L/2)\right]\,.
\end{multline}
In the limit $L\gg d$ this relation simplifies to

\begin{equation}\label{Eqs-approx-ee}
    \int\limits_{-L/2}^{L/2} {\rm d}y V_0e^{-|y|/d}f(x+y) \approx 2V_0d\sum_{n=0}^\infty d^{2n}f^{(2n)}(x)\,.
\end{equation}
Using the relation \eqref{Eqs-approx-ee} and substitution $z=2\pi
x/L$ we can rewrite the matrix elements

\begin{equation}
    V_{ij}=\frac{1}{2}\sum\limits_{\alpha,\beta,\gamma,\delta} a_{ij\alpha\beta\gamma\delta} V_{\alpha\beta\gamma\delta}
\end{equation}
with $a_{ij\alpha\beta\gamma\delta}$ being one of the values
$\{-1,0,1\}$ and with

\begin{multline}
    V_{\alpha\beta\gamma\delta}=\\
        V_0\!\!\int\limits_{-L/2}^{L/2}\!\!\! {\rm d}x\; \psi_{\alpha}^*(x)\psi_{\gamma}(x)\!\!
        \int\limits_{-L/2}^{L/2}\!\!\! {\rm d}y\; e^{-|y|/d}\,\psi_{\beta}^*(x+y)\psi_{\delta}(x+y)
\end{multline}
into the form

\begin{equation}
    V_{ij}\approx V_0d\frac{L}{2\pi}\sum_{n=0}^{\infty}\left(\frac{2\pi d}{L}\right)^{2n}A_{ijn}\,,
\end{equation}
where

\begin{equation}
    A_{ijn}=\sum\limits_{\alpha,\beta,\gamma,\delta} a_{ij\alpha\beta\gamma\delta}
        \int\limits_{-\pi}^{\pi}\!\!\!{\rm d}z\;
        \psi_\alpha^*(z)\psi_\gamma(z)\frac{\partial^{2n}}{\partial z^{2n}} [\psi_\beta^*(z)\psi_\delta(z)]\,.
\end{equation}
Here, the term with $n=0$ corresponds to the $\delta$-function
like electron-electron interaction. Due to the Pauli exclusion
principle, this term does not contribute to the total energy and
total current. The second term in the summation over $n$ is
proportional to $d^3$. If we parametrize the exponential
electron-electron interaction $V(y)=V_0e^{-|y|/d}$ by the
parameter $\alpha_{RG}$ and express $V_0$ from that $\alpha_{RG}$,
our matrix elements $V_{ij}$ exhibit two limiting cases. In the
case $4k_F^2 d^2 \ll 1$, where $V_0\sim \alpha_{RG}/d^3$, the
leading term in the summation over $n$ is a $d$-independent
constant. This is the reason why our many-body calculations give
for small enough $d$ the $d$-independent persistent current. In
the case $4k_F^2 d^2 \gg 1$, where $V_0\sim \alpha_{RG}/d$, the
summation involves the terms proportional to $d^2$, $d^4$, etc,
i.e., the matrix element is a complicated function of $d$. This is
the reason why our calculations for large $d$ give the persistent
current depending on two parameters, $d$ and $V_0$.

%


\begin{thebibliography}{00}
\bibitem{remark1}
The equation \eqref{Eqs-trancedent} can be found say in Ref.
\cite{Gogolin}, but written in a different form.
\bibitem{commentI}
In the local approximation \eqref{Eqs-Cohen} the pesistent current
reads $I = -\sum_k\left[ \frac{\partial\varepsilon_k}{\partial
\phi} -
\left\langle\psi_k|\frac{\partial}{\partial\phi}(U_H(x)+U_F(x))|\psi_k\right\rangle\right]$.
This formula can be derived \cite{Cohen-97} by inserting into
\eqref{Eqs-Vseob-Prud} the ground-state energy \eqref{HF ground
state energy} and by applying the 'almost closure relation'
$\sum_{n'} \psi_{n'}^*(x')\psi_{n'}(x) \simeq \delta(x-x')$.
\end{thebibliography}
\end{document}